\documentclass{aa501}
\usepackage{graphicx}
\usepackage[authoryear]{natbib}
\usepackage{psfig}

\begin{document}

\title{Extragalactic Globular Clusters in the Near-Infrared}

\subtitle{II. The Globular Clusters Systems of NGC~3115 and NGC~4365
  \thanks{Based on observations obtained at the European Southern
    Observatory, Chile (Observing Programme 63.N--0287).} 
   \thanks{Based
    on observations made with the NASA/ESA Hubble Space Telescope,
    obtained from the data archive at the Space Telescope Institute.
    STScI is operated by the association of Universities for Research in
    Astronomy, Inc. under the NASA contract NAS 5-26555.}}
 
   \author{Thomas H. Puzia \inst{1}, Stephen E. Zepf \inst{2,3}, Markus
     Kissler-Patig \inst{4}, Michael Hilker\inst{5}, Dante Minniti
     \inst{6}, \and Paul Goudfrooij \inst{7}}
      
   \offprints{Thomas H. Puzia, \email{puzia@usm.uni-muenchen.de}}

   \institute{Sternwarte der Ludwig-Maximilians-Universit\"at,
        Scheinerstr. 1, 81679 M\"unchen, Germany, \\
        \email{puzia@usm.uni-muenchen.de}
      \and 
      Department of Astronomy, Yale University, P.O. Box 208101,
        New Haven, CT 06520-8101, USA
      \and 
      Department of Physics and Astronomy, Michigan State University,
        East Lansing, MI 48824, USA, \\\email{zepf@pa.msu.edu}
      \and
      European Southern Observatory, 85749 Garching bei M\"unchen,
        Germany, \\\email{mkissler@eso.org}
      \and 
      Sternwarte der Universit\"at Bonn, Auf dem H\"ugel 71, 53121 Bonn,
        Germany, \\\email{mhilker@astro.uni-bonn.de}
      \and
      Departamento de Astronom\'\i a y Astrof\'\i sica,
        P.~Universidad Cat\'olica, Casilla 104, Santiago 22, Chile,
        \\\email{dante@astro.puc.cl}
      \and 
      Space Telescope Science Institute, 3700 San Martin Drive,
        Baltimore, MD 21218, USA, \\\email{goudfroo@stsci.edu}
        }

   \authorrunning{Puzia et al.}
   \titlerunning{Extragalactic Globular Clusters in the Near-Infrared II.} 
      
   \date{Received February, 2002; accepted June, 2002}
   
   \abstract{We combine near-infrared photometry obtained with the
     VLT/ISAAC instrument and archival HST/WFPC2 optical images to
     determine $VIK$ magnitudes and colours of globular clusters in
     two early-type galaxies, NGC~3115 and NGC~4365. The combination
     of near-IR and optical photometry provides a way to lift the
     age-metallicity degeneracy. For NGC 3115, the globular cluster
     colours reveal two major sub-populations, consistent with
     previous studies. By comparing the $V-I$, $V-K$ colours of the
     NGC~3115 globular clusters with stellar populations models, we
     find that the colour difference between the two $\ga10$ Gyr old
     major sub-populations is primarily due to a difference in
     metallicity. We find $\Delta$[Fe/H]$=1.0\pm0.3$ dex and the blue
     (metal-poor) and red (metal-rich) globular cluster
     sub-populations being coeval within 3 Gyr. In contrast to the
     NGC~3115 globular clusters, the globular cluster system in
     NGC~4365 exhibits a more complex age and metallicity
     structure. {\it We find a significant population of intermediate-age
     very metal-rich globular clusters} along with an old population
     of both metal-rich and metal-poor clusters. Specifically, we
     observe a large population of globular clusters with red $V-K$
     colours but intermediate $V-I$ colours, for which all current
     stellar population models give ages and metallicities in the
     range $\sim2-8$ Gyr and $\sim0.5Z_\odot- 3Z_\odot$,
     respectively. After 10 Gyr of passive evolution, the
     intermediate-age globular clusters in NGC~4365 will have colours
     which are consistent with the very metal-rich population of
     globular clusters in giant elliptical galaxies, such as M87. Our
     results for both globular cluster systems are consistent with
     previous age and metallicity studies of the diffuse galactic
     light. In addition to the major globular cluster populations in
     NGC~3115 and NGC~4365 we report on the detection of objects with
     extremely red colours ($V-K\ga3.8$ mag), whose nature could not
     ultimately be revealed with the present data.}
   
   \maketitle


\section{Introduction}
\label{ln:intro}
Globular cluster systems are useful tracers of galaxy evolution. They
consist of bright clusters, which are made of stars that share the same
age and chemical composition. Globular clusters can form during major
merger events in galaxies \citep[e.g.][etc.]{holtzman92, whitmore93,
  whitmore95, schweizer96}, but we also observe the formation of massive
star cluster in galaxies with a moderate star formation rate
\citep[e.g.][etc.]{oconnell94, barth95, oconnell95, brandl96, larsen99,
  hunter00}. In recent years the study of globular cluster systems
revealed the presence of globular cluster sub-populations which must
have formed in multiple formation epochs and/or mechanisms
\citep[see][for reviews]{ashman98, kisslerpatig00, vandenbergh00,
  harris01}.

In practice, for distant galaxies, we observe integrated properties of
their globular clusters, and would like to deduce their physical
properties, in particular their age and metallicity. Unfortunately, even
in systems where reddening can be taken as uniform, the optically
observed parameters suffer from the well known age-metallicity
degeneracy. Spectroscopy can overcome this infamous problem
\citep{jones95, worthey97, vazdekis99a} to a large extent, but is,
however, very time consuming to perform for hundreds of objects.
Photometry still represents the most efficient way to study an entire
globular cluster system.

There have been several attempts to solve this degeneracy for globular
cluster systems in early-type galaxies which employed optical colours
only \citep[e.g.][]{kisslerpatig97, whitmore97, kisslerpatig98, kundu99,
  puzia99} with various degrees of success. Previous studies used the
optical {\it colour difference and the turn-over magnitude difference}
between the major globular cluster sub-populations to derive an age and
metallicity difference. This method suffers from two main problems: 1)
It is based on the assumption that the mass functions are similar for
both sub-populations, so that luminosity differences directly reflect
differences in the mass-to-light ratios of the individual populations.
2) The peak colour and turn-over magnitude differences are small and
somewhat model dependent.

Alternatively one could derive mean age and metallicity differences
between globular cluster sub-populations from {\it colour-colour
  diagrams}, which does not depend on the globular-cluster mass
functions.

The combination of optical and near-infrared (near-IR) photometry can
largely reduce the age-metallicity degeneracy. This is because optical
to near-IR indices like $V-K$ are very sensitive to metallicity, but
only have modest age sensitivity \citep[hereafter Paper
I]{kisslerpatig00, puzia01, kisslerpatig02}. Physically, this technique
works because the $V$-band samples mainly the light of stars near the
turn-off, while the $K$-band is most sensitive to cooler stars on the
giant-branch in old stellar populations \citep[see e.g.][]{yi01}. While
the turn-off is mostly affected by age, the giant branch is primarily
sensitive to metallicity \cite[e.g.][]{saviane00}. Thus, $V-I$ and $V-K$
are affected similarly by age, but $V-K$ is much more sensitive to
metallicity (see Table \ref{tab:slopes} below). This allows the
age-metallicity degeneracy to be lifted by determining the location of a
simple stellar population in a plot of $V-I$ vs. $V-K$.

Such diagrams are powerful tools to study the age and the metallicity of
globular-cluster populations. Including near-IR passbands to our optical
data we can increase the sensitivity to metallicity differences by a
factor of two or more (see Paper I, and SSP models of e.g.
\citealt{bc00}).

\begin{table}[t!]
\centering
\caption[width=\textwidth]{Basic information on observed galaxies. The
  references are (1): \cite{RC3}, (2): \cite{schlegel98}, (3):
  \cite{buta95}, (4): \cite{frogel78}. (5): \cite{tonry01}. }
\label{tab:galdat}
\begin{tabular}{l r r l}
\hline
\noalign{\smallskip}
Parameter & NGC~3115 & NGC~4365 & Ref. \\
\noalign{\smallskip}
\hline
\noalign{\smallskip}
type               & S0             & E3              &(1) \\
RA  (J2000)        &10h 05m 14s     & 12h 24m 28s     &(1) \\
DEC (J2000) &$-07^{\rm o}$ 43' 07'' & $+07^{\rm o}$ 19' 03'' &(1)\\
$l$                &$247.78^{\rm o}$&$283.80^{\rm o}$ &(1)\\
$b$                &$36.78^{\rm o}$ &$ 69.18^{\rm o}$ &(1)\\
$B_{\rm T,0}$      & 9.74           & 10.49           &(1) \\
E$_{B-V}$          & 0.047          &  0.021          &(2)\\
$(B-V)_{\rm o}$    & 0.94           &  0.95           &(1)\\
$(V-I)_{\rm eff,o}$& 1.25           &  1.25           &(3)\\
$(V-K)_{\rm eff,o}$&$3.30\pm0.02$   & $3.29\pm0.1$    &(4)\\
$(m-M)_V$          &$29.93\pm0.09$  & $31.55\pm0.17$  &(5)\\
$M_V$              &$-21.13$        &$-22.01$         &(1),(5)\\
\noalign{\smallskip}
\hline
\end{tabular}
\end{table}

In this paper we study the globular cluster systems of NGC~3115 and
NGC~4365. NGC 3115 is an isolated galaxy, located at the very tip of the
southern extension of the Leo Group, with just one significant
accompanying nucleated dwarf elliptical galaxy \citep{puzia00a}. It
features a bimodal optical colour distribution of globular clusters
\citep{kundu98,gebhardt99,larsen01}.

NGC~4365 is a cluster giant elliptical galaxy located at the outer
edge of the Virgo Cluster. Optical photometry of the globular cluster
system revealed a broad but single-peak colour distribution with
little evidence for bimodality \citep{forbes96, gebhardt99, larsen01,
kundu01a}. \citeauthor{surma95} (\citeyear{surma95}, see also
\citealt{davies01}) detected a decoupled, counter-rotating core, which
consists of a younger and more metal-rich stellar population than the
rest of the galaxy. Such decoupled cores are believed to be created in
major merger events. All relevant galaxy properties are summarized in
Table \ref{tab:galdat}.

The major goal of this paper is to derive ages and metallicities of
globular clusters from a comparison of optical/near-IR colours with
several SSP models. Putting this results into a larger context,
including previous findings from galactic integrated light studies, we
can further constrain formation scenarios for both galaxies.

The outline of this paper reads as follows. In Section 2 we describe the
data reduction and calibration of our new near-IR VLT/ISAAC and the
archival optical HST/WFPC2 data. The main results are presented in
Section 3 followed by a discussion in Section 4. All major findings are
summarized in Section 5.

\section{The Photometric Data}
\subsection{New VLT/ISAAC Near--Infrared Data}
\subsubsection{Basic Reduction}
Deep images in $K_s$\footnote{The $K_s$ ({\it K short}) filter is a
slightly modified $K$-band filter with the wavelength range
$2.03-2.30\mu$m. Contrary to the $K$ filter the transmission curve
drops before the CO-absorption (at $\sim2.4\mu$m) band making the
$K_s$ filter insensitive to CO abundance variations. Moreover, the sky
background is reduced relative to the standard $K$-band due to the
short-wavelength cut-off.} were obtained in service mode (program
63.N--0287) with the Near-Infrared Spectrometer And Array Camera
\citep[ISAAC, see][]{moorwood98} attached to the Unit Telescope 1
(Antu) of the European Southern Observatory's Very Large Telescope
(VLT). ISAAC is equipped with a Rockwell Hawaii 1024$\times$1024
pixels Hg:Cd:Te array. The pixel scale is 0.147\arcsec /pixel and the
field-of-view 2.5\arcmin $\times$ 2.5\arcmin\ on the sky. In the
following we will always refer to the $K_s$ filter as $K$. The ISAAC
field was oriented such as to match the WFPC2 field of view.

The NGC~3115 data were obtained during the nights of April 4th to 10th
1999 while the NGC~4365 data were taken during the nights of April 9th
and 10th and additionally on the night of June 2nd 1999. The exposures
were split into $O-SSSSS-O$ sequences, where $S$ was a dithered sky
exposure comprising of 5 single exposures and $O$ was a single on-target
observation. Each sky image was exposed with a detector integration time
(DIT) of 10 seconds and two such images were averaged in the read-out
electronics (NDIT=2). Thus, the total integration time for a sky image
is 20 seconds. An object image has a total integration time (DIT
$\times$ NDIT) of $10 \times 10 = 100$ seconds. Since the night-sky
luminosity changes on short time scales the crucial part in each near-IR
data reduction is the proper treatment of sky subtraction. Our
observation pattern allowed us to create mean sky images, which have
been cleaned of stars prior to combination to a {\it mastersky} image.
The data of each night have been processed individually. From each raw
object image the {\it mastersky} was subtracted giving a
sky/bias-subtracted object image. Those sky/bias-subtracted images were
divided by a normalized {\it masterskyflat}, which is provided by the
ESO quality control group. Eventually, all images were aligned using
15-20 objects common to all frames and averaged to give the final
cleaned object image. The FWHM of the stellar PSF in the final $K$ band
image is $\sim$0.6 \arcsec\ for NGC~3115 and NGC~4365. The total
exposure time for NGC~3115 is 15500 sec, and 9500 sec for NGC~4365.

\subsubsection{Photometric Calibration}
\label{ln:photcal}
For the photometric calibration of the NGC~3115 and NGC~4365 data set
11 and 9 near-IR standard stars have been taken throughout the nights
\citep{persson98}. Each standard star was imaged five times, each
exposure having the star centered at a different position on the chip.
Standard-star photometry was performed with the photometry tool
SExtractor v2.1.6 \citep{bertin96} in an aperture of 30 pixels of
diameter ($7.4\times$seeing). The corrections to an infinite-diameter
aperture were found to be negligible in a subsequent curve-of-growth
analysis. After applying a $\kappa-\sigma$ clipping to all the
measurements outliers have been deleted leaving 40 data points for the
calibration of each data set. Due to the lack of sufficient spread in
airmass among the standard-star measurements, we use the airmass term
determined in the standard calibration of the ISAAC instrument for
April 1999. Assuming that the colour terms are negligible in the
near-infrared\footnote{The theoretical value for the $J-K_s$ colour
term was determined by the data quality control group of the Paranal
observatory being close to zero (see ISAAC Data Reduction Guide
1.5). Since no $J$-band data of NGC~3115 and NGC~4365 are available
and the optical and near-IR photometric systems of Johnson and Persson
do not overlap, i.e. they have no stars in common, it is not possible
to include a $V-K$ colour term into the calibration.}, we find the
following calibration relations for the photometric nights serving as
reference
\begin{eqnarray}
      K_{\mathrm{3115}} & = & k_{\mathrm{inst}} + 23.56(\pm0.018) -
      0.05(\pm0.005)\chi \\   
      K_{\mathrm{4365}} & = & k_{\mathrm{inst}} + 23.75(\pm0.015) -
      0.05(\pm0.005)\chi
\end{eqnarray}
where $K_{\rm galaxy}$ is the calibrated magnitude, $k_{\rm inst}$ is
the instrumental magnitude normalised to 1 second, and $\chi$ the
airmass. The error of the zero points includes photometric errors of
each single standard star measurement and the errors of the
curve-of-growth analysis. The error of the airmass term is a estimate
from the variations in airmass of all single exposures.

After correcting all nights to a common zero point (obtained from a
reference photometric night) we traced the calibrated magnitude of two
isolated bright stars over all nights. We find an upper-limit
1-$\sigma$ scatter being a few hundredths of a magnitude. Thus, we
conservatively estimate the true photometric uncertainty, mainly
driven by the strongly varying sky background, to be of the order of
$\leq0.03$ mag.

Finally, all magnitudes were corrected for Galactic foreground reddening
using the reddening values of Table \ref{tab:galdat} and the extinction
curves of \cite{cardelli89}. The corrections for NGC~3115 and NGC~4365
are $A_K=0.017$ mag and $A_K=0.008$ mag.

\subsection{Archival HST/WFPC2 Optical Data}
\label{ln:hstcal}
NGC~3115 was observed with HST + WFPC2 under program GO.5512, with the
PC centered on the nucleus of the galaxy. The total exposure times of
the combined images are $3\times350$ sec in F555W, and $3\times350$ sec
in F814W. NGC~4365 has been imaged with WFPC2 under program GO.5920 in
F555W and F814W filters with 2200 sec and 2300 sec of total exposure
time, respectively. The HST images were reduced and calibrated following
for most parts the procedure as described in \cite{puzia99}. Briefly
recapitulating, all sub-exposures were matched by whole-pixel shifts and
stacked using a cosmic-ray cleaning routine (crrej) within
IRAF\footnote{IRAF is distributed by the National Optical Astronomy
  Observatories, which are operated by the Association of Universities
  for Research in Astronomy, Inc., under cooperative agreement with the
  National Science Foundation.}. The object finding was performed on a
master image which was created from all available exposures of galaxy to
allow reliable cosmic-ray cleaning. None of the images suffers from
crowding. Therefore, all magnitudes were measured with the SExtractor
tool using a 8 and 4-pixel-diameter aperture for the WF and PC chip,
respectively. Aperture corrections to the Holtzman standard aperture
\citep[0.5\arcsec, see][]{holtzman95} were determined in a
curve-of-growth analysis. Only objects within the expected colour range
for globular clusters were used (see Sect.~\ref{ln:selection}). The
correction terms can be found in Table \ref{tab:apcor}. Instrumental
magnitudes were then transformed to the Johnson $V$ and $I$ magnitudes
according to the prescription given in \citeauthor{holtzman95}. All
magnitudes were reddening corrected using the reddening values as listed
in Table \ref{tab:galdat}. Together with the extinction curves of
\cite{cardelli89} we obtain $A_V=0.157$ and $A_I=0.092$ in the direction
of NGC~3115 and $A_V=0.070$ and $A_I=0.041$ for NGC~4365.

\begin{table}[!t]
\centering
\caption{Aperture corrections for WFPC2 photometry. The given
  terms refer to the correction from an 8/4-pixel-diameter aperture (PC/WF
  chips) to the standard Holtzman 0.5\arcsec\ aperture. \# indicates the
  number of objects which were used to determine the correction.}
\label{tab:apcor}
\begin{tabular}{llccccr}
\hline
\noalign{\smallskip}
 & & PC1 & WF2 & WF3 & WF4 & $\langle$WF$\rangle$ \\
\noalign{\smallskip}
\hline
\noalign{\smallskip}
NGC~3115 & V & -0.152 & -0.296 & -0.385 & -0.323 & -0.323 \\
         & I & -0.236 & -0.370 & -0.412 & -0.330 & -0.362 \\
         &\# & 6      & 25     & 11     & 24     &        \\
\noalign{\smallskip}
\hline
\noalign{\smallskip}
NGC~4365 & V & -0.065 & -0.134 & -0.159 & -0.127 & -0.141 \\
         & I & -0.133 & -0.205 & -0.242 & -0.219 & -0.225 \\
         &\# & 28     & 66     & 39     & 67     &        \\
\noalign{\smallskip}
\hline
\end{tabular}
\end{table}

In the case of NGC~3115 excellent agreement was found between our data
set and the previous studies of \cite{kundu98} and \cite{gebhardt99}.
The systematic offset between the \citeauthor{kundu98} photometry and
ours is $0.04\pm0.11$ mag in $V_{\rm F555W}$ and $-0.01\pm0.11$ in
$I_{\rm F814W}$. We find also good agreement for NGC~4365 between the
photometry of \cite{gebhardt99} and our data \citep[but see
also][]{puzia00err}. In $V_{\rm F555W}$ we measure an offset
$0.04\pm0.07$ mag and $0.07\pm0.05$ mag for the $I_{\rm F814W}$
filter.

\subsection{Selection of Globular Cluster Candidates}
\label{ln:selection}
\begin{figure}
   \centering
   \includegraphics[width=8.5cm]{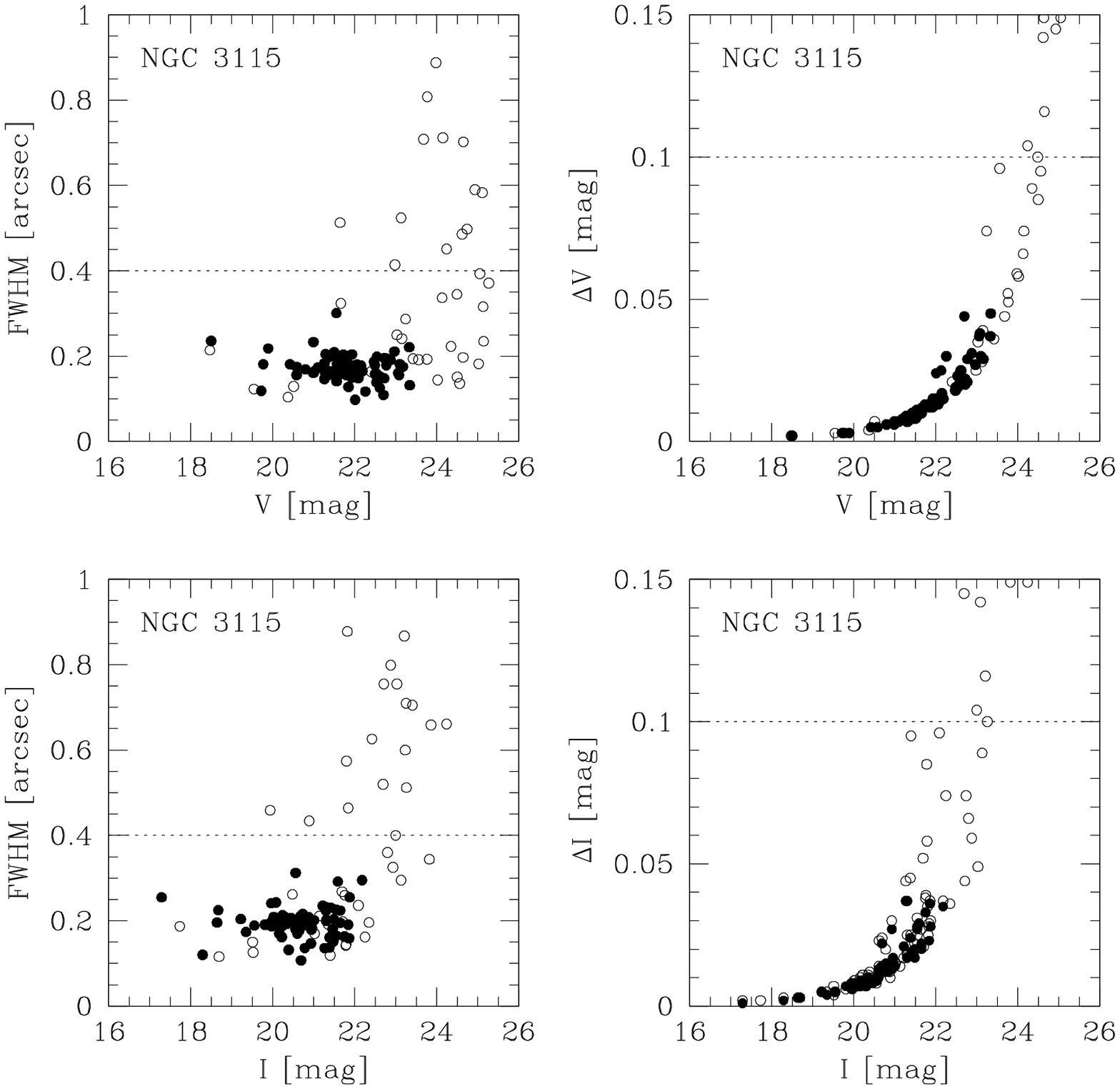}
   \includegraphics[width=8.5cm]{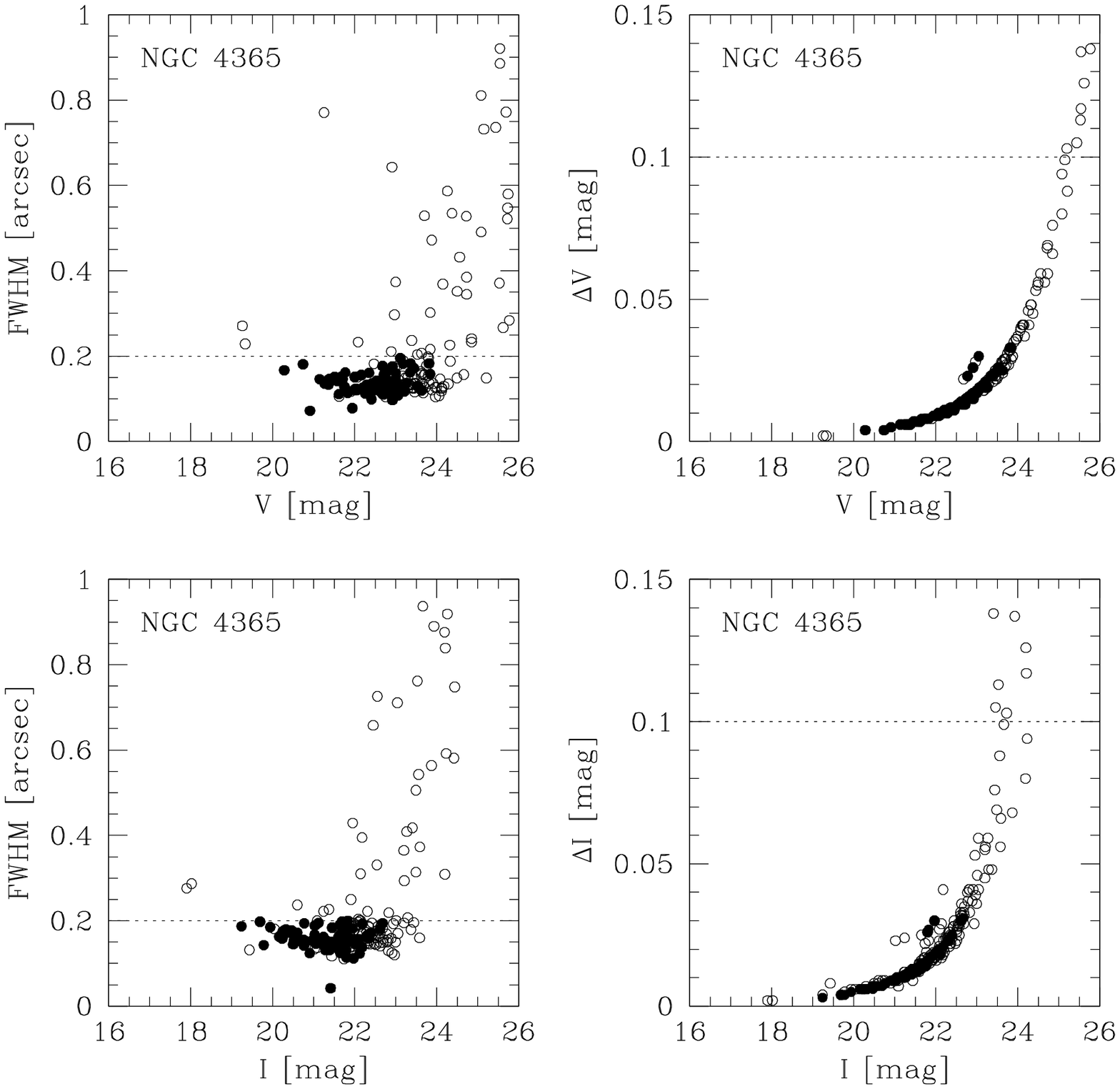}
      \caption{Selection parameter as a function of $V_{\rm F555W}$
        and $I_{F814W}$ magnitude. Plotted here are the full width at
        half maximum (FWHM) of the point-spread function (PSF) and the
        photometric error, as derived on HST images. Data points which
        were rejected during the selection are marked as open circles.
        Dotted lines indicated the selection boundaries.}
         \label{ps:cuts}
\end{figure}

For object selection the near-infrared ISAAC data were combined with
the archival optical HST data. Note that the WFPC2 field size of
2.66\arcmin$\times$2.66\arcmin\ is well matched to the ISAAC field of
view (2.5\arcmin$\times$2.5\arcmin ). After transformation of HST
coordinates to the ISAAC frame we required the matches being better
than 0.74\arcsec\ ($\sim$ seeing). Before selecting for likely GCs,
the matched lists contain 110 and 186 objects with $VIK$ photometry
for NGC~3115 and NGC~4365, respectively. To get rid of extended
sources and noisy detections, we selected GC with cuts in FWHM$_{\rm
WFPC2}$ and photometric error, as measured on the optical HST data.
Figure \ref{ps:cuts} illustrates the constraints which define our
globular cluster candidates. For NGC~3115, all objects with a
FWHM$_{\rm WFPC2}$ larger than 0.4\arcsec\ and for NGC~4365 larger
than 0.2\arcsec\ were rejected. Both FWHM$_{\rm WFPC2}$ values roughly
correspond to 20 pc at the distance of 9.7 Mpc and 20.4 Mpc for
NGC~3115 and NGC~4365, respectively \citep[see][]{tonry01}. To
determine the real sizes of clusters which are rejected by the
FWHM$_{\rm WFPC2}$ cut we convolve the stellar PSF in each image with
a series of King profiles \citep{king62} with a mean concentration
parameter $c=\log(r_t/r_c)=1.55$, which was determined from the Milky
Way globular cluster system \citep{harris96}. The tests yield a
maximum allowed core radius $r_c\approx5.6$ pc and 4.9 pc, for
NGC~3115 and NGC~4365 globular clusters, to be retained in the sample.
An inspection of the core radii of Milky Way globular clusters shows
that 82\% and 79\% have core radii smaller than the two respective
maximum allowed sizes mentioned above. Unless the size distribution of
NGC~3115 and NGC~4365 globular clusters differs significantly, which
seems unlikely given the similarity in sizes observed for many other
early-type galaxies \citep[e.g.][]{kundu98, larsen01}, we will miss
only the extremely large globular clusters by applying the FWHM$_{\rm
WFPC2}$ cuts.

Furthermore, only objects with photometric errors in the HST photometry
of less than 0.1 mag in $V$ and $I$ were retained. After the selection
procedure the NGC~3115 list contains 87 candidates while the NGC~4365
list comprises 136 globular cluster candidates\footnote{The final lists
  are available in electronic form from THP.}.

Background contamination is unlikely to make a significant contribution
to the selected samples. \cite{puzia99} have shown that with similar
selection criteria the background contamination is less than $5\%$.

\subsection{Completeness}
\label{ln:completeness}
The HST data for NGC~3115 is complete down to $V_{F555W}\la26$ and
$I_{F814W}\la25$. For NGC~4365 the photometry reaches $V_{F555W}\la26.5$
and $I_{F814W}\la25.5$. Clearly, the limiting photometric passband is
the $K$ filter for both our galaxies and it will define the completeness
of the data. Artificial star experiments with 1000 cycles, each
including 100 artificial objects into the $K$ images, yield a limiting
magnitude (50\% completeness) of $K\la20.0$ and $K\la20.25$ for NGC~3115
and NGC~4365, respectively.

Despite the much longer exposure time for NGC~3115 (15500 sec) the final
$K$-band limiting magnitude is about the same as for NGC~4365 (9500
sec). Purely from the ratio of exposure times one would expect the
NGC~3115 photometry being $0.53$ mag deeper than for NGC~4365.  However,
we find the NGC~4365 photometry is $0.25$ mag deeper which adds up to a
total difference of $\sim0.78$ mag or a factor $\sim1.4$ in S/N.

\subsubsection{Surface Brightness Fluctuations}
This discrepancy cannot be explained by a higher noise due to a
difference in surface brightness between the halos of NGC 3115 and NGC
4365, since they are measured to be the same within 0.1 mag/arcsec$^2$.
However, a factor contributing to the lower detection limit in the NGC
3115 field are the larger $K$-band surface brightness fluctuations
(SBFs) \citep{tonry88}:

In addition to the Poisson noise of the flux, each resolutions element
contains SBFs which are the Poissonian fluctuations of the sampled
number of stars. If the resolution element samples $n$ stars of the
mean flux $\bar{F}$ then the fluctuation of the flux will be
$\sigma_F=\bar{F}\sqrt{n}$. Hence, putting a galaxy twice as far away
from the observer will keep its surface brightness constant (one
resolution element samples now 4 times more stars which have 4 times
lower flux) but lower the SBFs by a factor 2. The SBF amplitude is
proportional to the distance $\sigma_F\propto D^{-1}$. Both NGC~3115
and NGC~4365 have roughly the same $K$-band surface brightness,
$\mu_{\rm 3115}=12.5$ and $\mu_{\rm 4365}=12.6$ mag arcsec$^{-2}$, at
radii including the majority of our globular cluster candidates.
However, since NGC~3115 is roughly two times closer than NGC~4365
\citep{tonry01}, the ratio of SBF amplitudes, $\sigma_F\propto$ S/N,
will be $\sim$ 2 (compared to $\sim 1.4$ measured). Thus, relaxing the
implicit assumption of equal stellar populations, the difference in
the SBF amplitude can explain the difference in the limiting
magnitudes.

\subsubsection{Colour Completeness}
Regardless of the limiting magnitude, our data samples are biased
toward red globular clusters. That is, the red (metal-rich)
sub-population is more complete than the blue (metal-poor) one at a
given magnitude threshold (see Fig.~\ref{ps:cmd}). To quantify the
biasing factor we assume that the globular cluster luminosity
functions (GCLFs) of both sub-populations have the same shape (same
mass function) and the same dispersion (same slopes of the mass
function). Both globular cluster sub-populations are assumed to have
the same M/L. Previous studies of the globular clusters systems of
elliptical and S0 galaxies suggest that this assumption is a
reasonable initial approximation, although in detail there is some
evidence for modest differences in the M/L between the metal-poor and
metal-rich sub-populations of well-studied cluster systems
\citep[see][for a discussion on NGC~4472 and M~87]{puzia99, kundu99}.

The turn-over magnitude of the NGC~3115 GCLF is $V_{\rm TO}=
22.6\pm0.2$ for both blue and red cluster candidates
\citep{larsen01}. Concerning the mean $V-K$ colour for the blue
($2.3\pm0.1$ mag, see Sect.~\ref{ln:coldistr} and Table~\ref{tab:kmm})
and for the red clusters ($3.0\pm0.1$ mag) the expected turn-over
magnitudes in $K$ are $K_{\rm TO,blue}\approx 20.3$ and $K_{\rm
TO,red}\approx 19.6$ mag. According to the completeness limit of the
NGC~3115 data ($K_{\rm lim}\approx20.0$ mag) and assuming a GCLF
dispersion $\sigma_{\rm GCLF}=1.3$, we sample 41\% of the blue and
62\% of the red sub-population. The biasing factor for the NGC~3115
globular cluster system is thus $\sim1.5$. That is, we detect
$\sim1.5$ times more red clusters than blue ones. In the case of
NGC~4365 we detect 13\% of blue and 23\% of red clusters assuming
$V_{\rm TO}=24.37\pm0.16$ mag along with $\sigma_{\rm GCLF}=1.22$ mag
\citep{larsen01} and $V-K=2.76$ and 3.23 mag for blue and red
clusters, respectively. The biasing factor is 1.8. If we assume a
single peak colour distribution with a mean $V-K\approx2.95$ (see
Sect.~\ref{ln:coldistr}) and a turn-over magnitude independent of
colour we sample 17\% of the GCLF in NGC~4365.

\subsubsection{Spatial Completeness}
\begin{figure*}
   \centering
   \includegraphics[width=8.5cm]{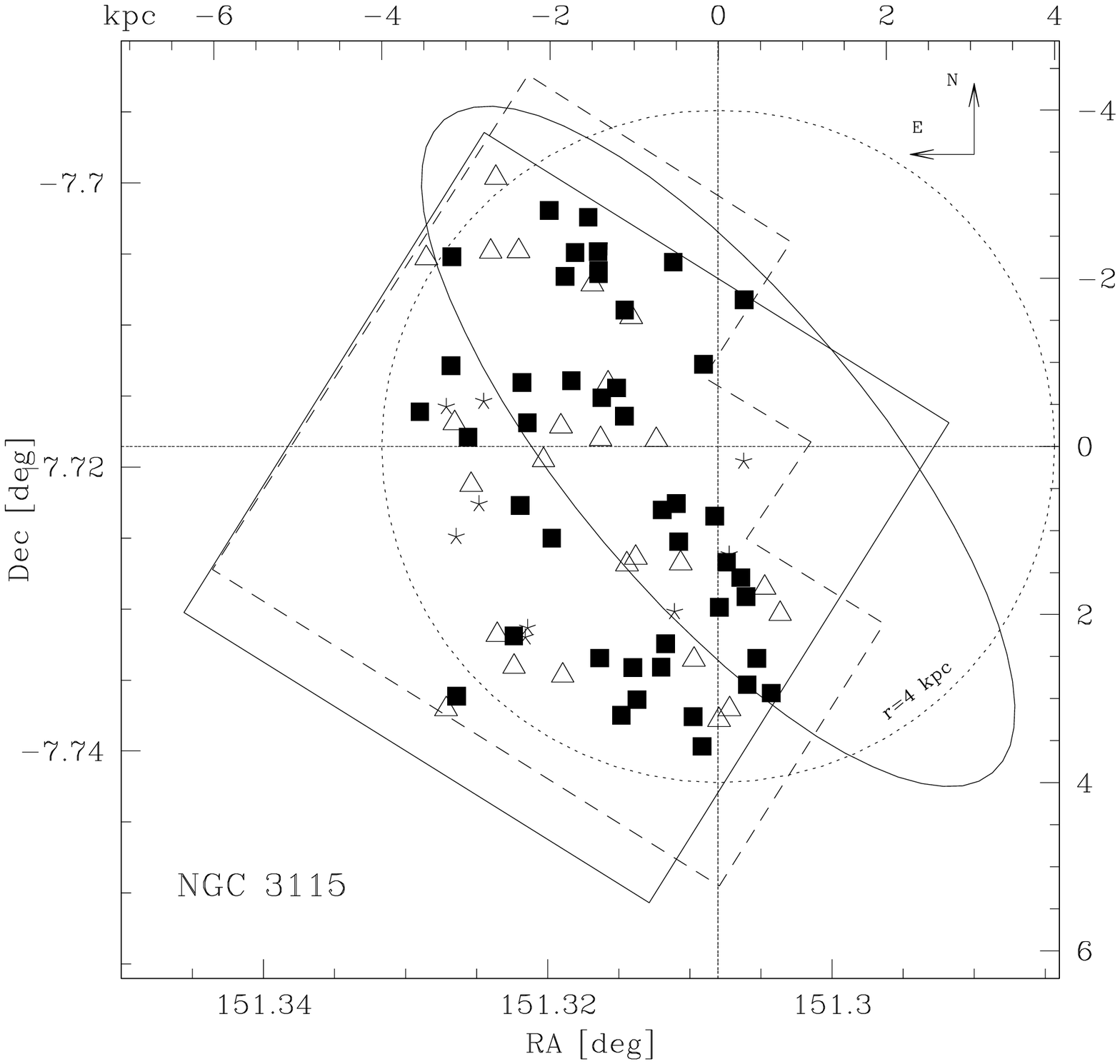}
   \includegraphics[width=8.5cm]{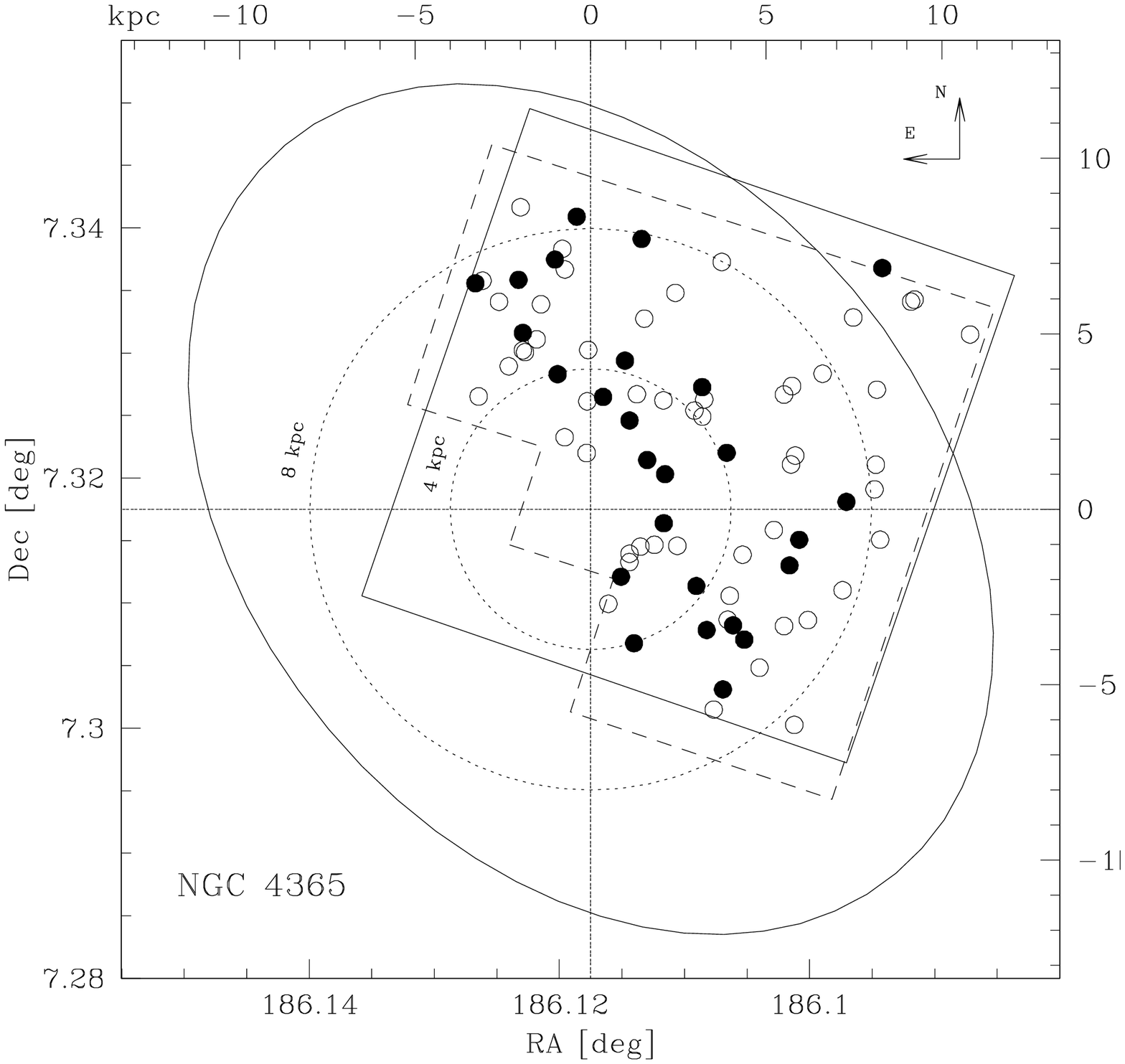}
      \caption{Field of view for the NGC~3115 (left panel) and NGC~4365
        (right panel) data on the sky. The solid square shows the
        effective aperture of the ISAAC data. The HST aperture is
        indicated as a dashed polygon. Solid ellipses indicate the area
        encompassing half of the integrated light of the galaxy. For
        NGC~3115, we split the sample at $V-I=1.05$ and plot the blue
        sub-population as open triangles and the red clusters as filled
        squares. Stars mark the position of extremely-red objects in the
        colour range $3.7< V-K < 4.3$, which meet our completeness but
        not the size selection criteria (all extremely-red objects lie
        above the dashed line in Fig.~\ref{ps:cmd}). For NGC~4365, solid
        circles show the positions of globular clusters with
        intermediate ages, whereas the remainder globular cluster
        candidates are marked by open circles.}
         \label{ps:fov}
\end{figure*}

Both our globular cluster samples are spatially incomplete (see
Fig.~\ref{ps:fov}). Moreover, the different spatial distribution of blue
and red globular clusters also influences the completeness. In NGC~3115
we sample the inner 4 kpc of the globular cluster system. The spatial
coverage of NGC~4365 globular clusters is more complete. Our field of
view covers clusters with galactocentric distances up to $\sim10$ kpc.
For NGC~3115, \cite{kundu98} report that the red globular clusters
follow the galactic light profile while the blue clusters rather reside
in the halo. Thus, independent of completeness biases, both samples have
to be considered as {\it spatially} biased in favour of red globular
clusters. The spatial distribution of globular cluster candidates is
shown in Figure \ref{ps:fov}, along with the location of the WFPC2 and
ISAAC apertures.

\section{Results}
\subsection{Colour Distributions and Colour-Magnitude Diagrams}
\label{ln:coldistr}
We use the KMM code \citep{ashman94} to test for bimodality of the
colour distributions. The code calculates the likelihood $P_{\rm KMM}$
whether or not a two-Gaussian fit is preferred over a single-Gaussian
fit and determines the positions of the modes. Higher $P_{\rm KMM}$
values indicate that a single-peak distribution is more likely. In
addition to the cuts in FWHM$_{\rm WFPC2}$ and photometric error (see
Sect.~\ref{ln:selection}), we select globular cluster candidates by
their $V-K$ colour in the range $1.7 < (V-K)< 3.7$ and cut the samples
at the limiting $K$ magnitude (NGC~3115: $K\la20.0$, and NGC~4365:
$K\la20.25$). According to simple stellar population models (see e.g.
Fig.~\ref{ps:colcol}) this colour range selects globular clusters with
metallicities $-2.5\la$[Fe/H]$\la+0.4$ dex and with a wide range of
ages. The input lists contain now 69 and 78 objects for NGC~3115 and
NGC~4365, respectively.

KMM returns for NGC~3115 a likelihood of $P_{\rm KMM}=0.53$ with peaks
at $V-K=2.31\pm0.06$ and $3.01\pm0.04$. Former studies found a clear
bimodality in the $V-I$ colour distribution of globular clusters in
NGC~3115. \cite{kundu98} find peaks at $V-I=0.96$ and $1.17$ and the
most likely separation of the two sub-populations being at $(V-I)_{\rm
  dip}=1.06$. \cite{gebhardt99}, who first calculated the likelihood of
a dip in a supposedly bimodal colour distribution, give $(V-I)_{\rm
  mean}=1.03\pm0.03$ with a dip likelihood $P_{\rm dip}=0.71$.
\cite{larsen01} find peaks\footnote{The errors of the colour peaks were
  kindly provided by S\o ren Larsen.} at $V-I=0.92\pm0.03$ and
$1.15\pm0.03$ and a dip likelihood $P_{\rm dip}=0.96$.

As for NGC~4365, KMM gives $P_{\rm KMM}=0.94$ with peaks at
$V-K=2.76\pm0.04$ and $3.23\pm0.04$. In NGC~4365 \cite{gebhardt99} find
a hint for bimodality in their $V-I$ colour distribution. The mean
colour and the bimodality significance of their distribution are
$(V-I)_{\rm mean}=1.09\pm 0.02$ and $P_{\rm dip}=0.76$. \cite{larsen01}
measure $V-I=0.98\pm0.03$ and $1.19\pm0.03$ as peak colours and a dip
likelihood $P_{\rm dip}=0.033$. These results show that there are some
galaxies for which extant colour distributions do not give a clear
result when statistical tests are applied to search for bimodality. No
clear multi-modality is detectable in either colour distribution
\citep[for $V-I$ see][]{gebhardt99, larsen01, kundu01a}. The globular
clusters show a continuous colour distribution.

\begin{figure*}
   \centering
   \includegraphics[width=10.5cm]{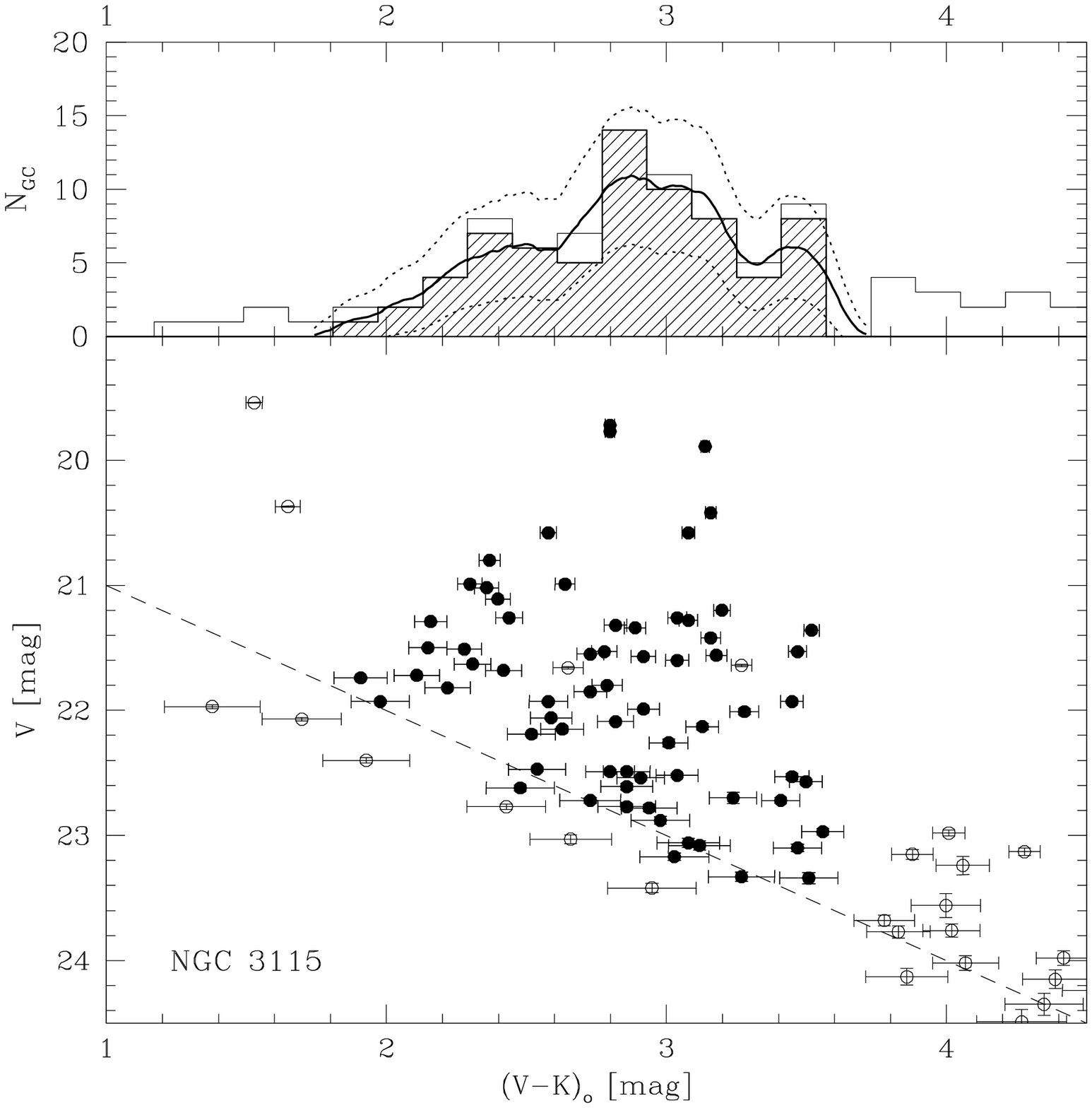}
   \includegraphics[width=10.5cm]{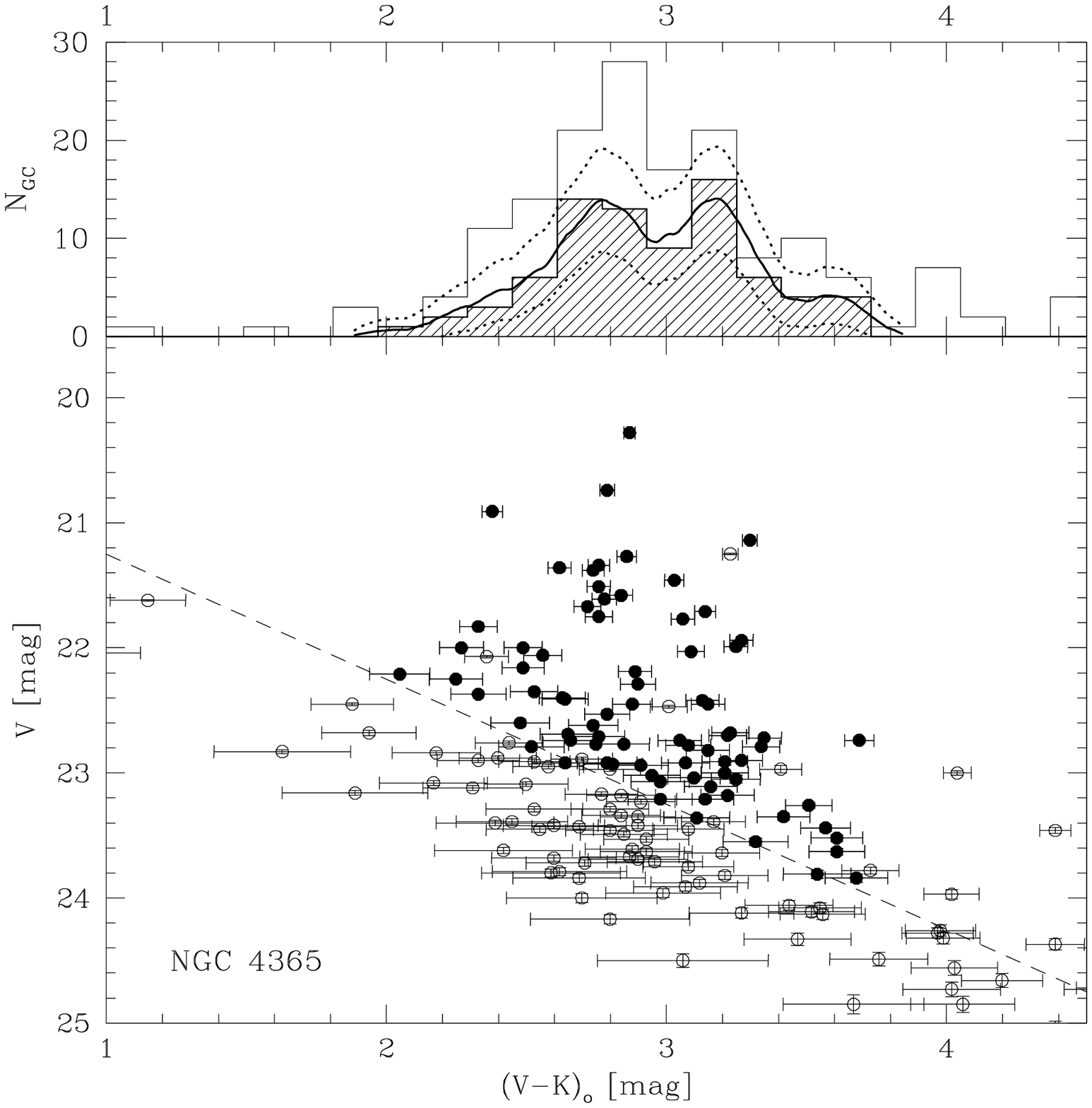}
      \caption{$V$ vs. $V-K$ colour-magnitude diagrams for NGC~3115
        (upper panel) and NGC~4365 (lower panel). The bottom of each
        panel shows the CMD of the associated globular cluster system.
        Filled symbols are selected GC while open symbols indicate data,
        which were rejected by our selection criteria (i.e., FWHM$_{\rm
          WFPC2}$, photometric error, and colour). The dashed line shows
        the 50\% completeness limit of the photometry as constrained by
        the near-IR photometry. The errors include photometric errors
        only. The top of each panel shows the $V-K$ histogram of all
        detected sources (solid histogram) and the selected GC sample
        (shaded histogram). A probability-density estimate
        \citep{silverman86, puzia00a} of the selected unbinned data is
        indicated (thick line) along with its 1$\sigma$ uncertainty
        limits (dotted lines).}
         \label{ps:cmd}
\end{figure*}

The errors of $V-K$ peak colours in our KMM analysis have been
determined by varying the input parameter of the KMM code, which are
the initial colour for both peaks, their covariance, and whether or
not the two peaks have the same dispersion. All values are summarized
in Table \ref{tab:kmm}. Note, that the driving parameters for
detecting bimodalities with KMM are the peak locations, the
dispersions, and the sample size. With smaller samples the $P_{\rm
KMM}$ parameter becomes less meaningful \citep{ashman94}. The
bimodality test shows that only the colour distribution of NGC~3115 is
likely to be not unimodal. Contrary, NGC~4365 has a globular cluster
colour distribution, which is well represented by a single
Gaussian. See Figure \ref{ps:cmd} for $(V-K)$ histograms.

\begin{table}[!t]
\centering
\caption{KMM mean colours of the globular cluster sub-populations in NGC~3115
  and NGC~4365.}
\label{tab:kmm}
\begin{tabular}{llcc}
\hline
\noalign{\smallskip}
 & & $V-K$ & $V-I^{\mathrm{a}}$ \\
\noalign{\smallskip}
\hline
\noalign{\smallskip}
NGC~3115 & blue & $2.31\pm0.06$ & $0.922\pm0.03$ \\
         & red  & $3.01\pm0.04$ & $1.153\pm0.03$ \\
\noalign{\smallskip}
\hline
\noalign{\smallskip}
NGC~4365 & blue & $2.76\pm0.04$ & $0.981\pm0.03$ \\
         & red  & $3.23\pm0.04$ & $1.185\pm0.03$ \\
\noalign{\smallskip}
\hline
\end{tabular}
\begin{list}{}{}
\item[$^{\mathrm{a}}$] values for the blue and red sub-populations
  adopted from \cite{larsen01}.
\end{list}
\end{table}

A metallicity calibration of $V-K$ using Milky Way and M31 globular
clusters with $E_{B-V}<0.27$ is derived in Paper I
$${\rm [Fe/H]} = -5.52(\pm0.26)+1.82(\pm0.11) \cdot (V-K)$$
with a rms
of 0.29 dex. With this calibration we calculate metallicities from peak
colours which were obtained in the KMM analysis. The globular cluster
sub-populations in NGC~3115 have then metallicities
[Fe/H]$=-1.32\pm0.11$ and $-0.04\pm0.07$ dex. Transforming the KMM peaks
of the NGC~4365 system yields [Fe/H]$=-0.50\pm0.07$ and $0.36\pm0.07$
dex. The median $V-K$ colours of the globular cluster systems of
NGC~3115 and NGC~4365 are equivalent to [Fe/H]$=-0.15\pm0.07$ and
$-0.33\pm0.05$ dex, respectively. The errors include the uncertainty in
the peak colours only. Note, that the calibration was established with
Milky Way and M31 globular clusters, which have very few clusters at
higher metallicities (i.e. at [Fe/H]$\ga-0.4$ dex). Consequently, all
derived values with metallicities around solar and higher may suffer
from uncertainties due to extrapolations.

\label{ln:kmm}

In Figure \ref{ps:cmd} (bottom panels) we show $V$ vs. $V-K$
colour-magnitude diagrams (CMDs). Together with the $V-K$ histograms in
Fig.~\ref{ps:cmd} (top panels) the CMDs of NGC~3115 and NGC~4365 show no
clear multi-modality due to the limited sample size. The dashed lines
show the 50\% completeness limits and make the selective completeness
obvious.

The CMDs also show the data which were rejected by our object selection
criteria (FWHM$_{\rm WFPC2}$, photometric error, and $V-K$), which are
shown as open circles. For NGC~3115, this includes two bright and blue
likely foreground stars. Using the galactic stellar population
model\footnote{The model predictions were calculated using the code at
  www.obs-besancon.fr/www/modele/modele\_ang.html} of \cite{robin96}, we
expect zero stars. In the case of NGC~4365 we find no blue contaminating
objects, in agreement with the model predictions.

\subsection{Extremely Red Objects}
\begin{figure}
   \centering
   \includegraphics[width=8.5cm]{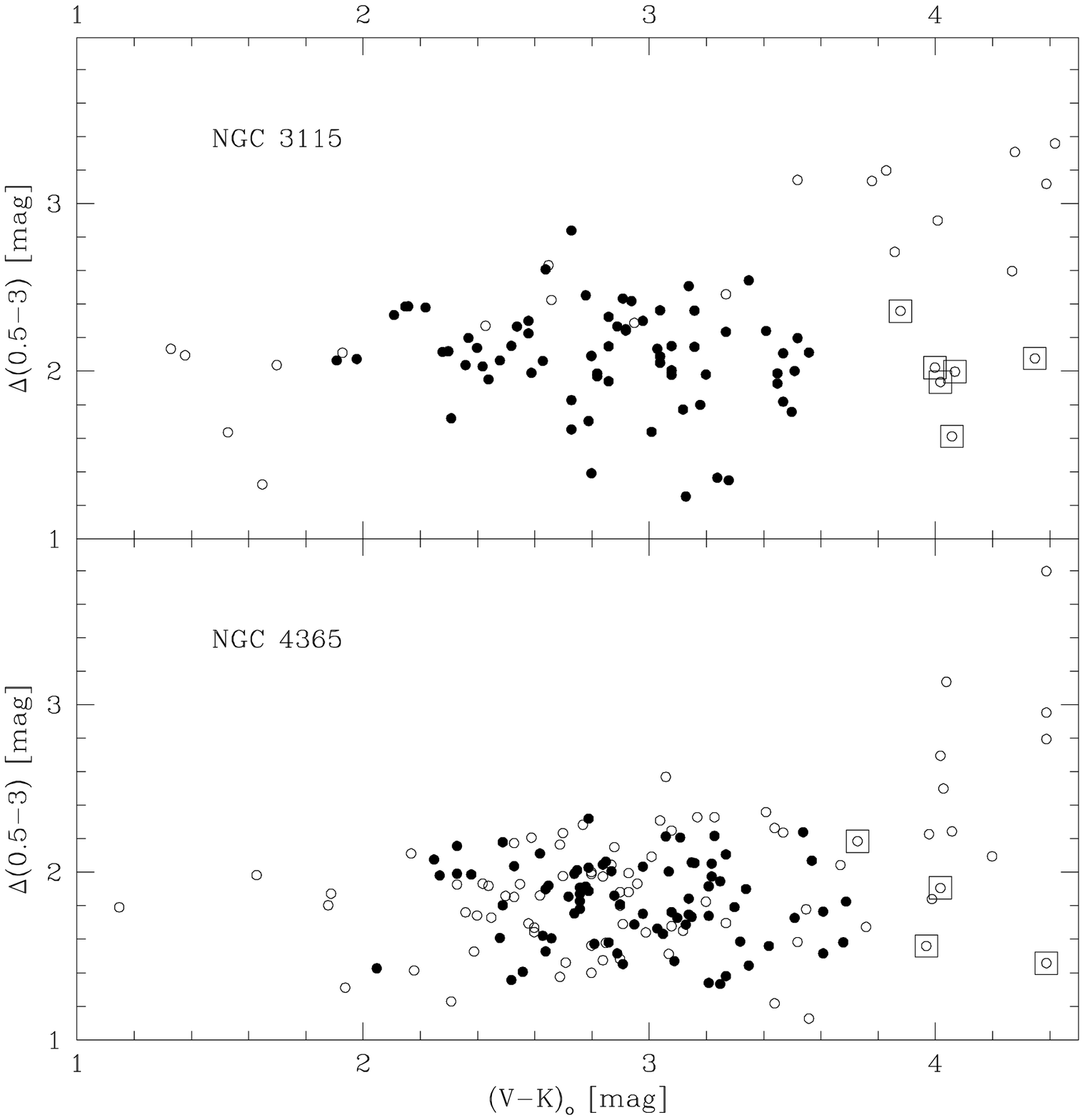}
      \caption{Relative object sizes as a function of $V-K$ colour for
        NGC~3115 (top panel) and NGC~4365 (lower panel). Solid points
        show the data after application of the selection criteria
        (FWHM$_{\rm WFPC2}$, photometric error, and $V-K$ colour)
        while open circles show all data. Open squares show data with
        red colours which would pass the selection criteria if the
        $V-K$ selection were dropped (i.e. point-like very red
        objects). Note, that there is no trend of size vs. colour
        seen in either galaxy, contrary to the results of
        \citeauthor{larsen01} who find that red globular clusters in
        near-by elliptical galaxies are on average 20\% smaller than
        blue ones. However, this correlation might be camouflaged by
        the colour bias of our sample.}
        \label{ps:sizes}
\end{figure}

At the red end of the NGC~3115 and NGC~4365 CMD (Fig.~\ref{ps:cmd}) we
find a dozen objects with $V-K\approx 4.0$ mag, which were rejected by
the FWHM$_{\rm WFPC2}$, completeness, and photometric-error cuts. These
objects show no significant clustering and are evenly distributed on the
sky. Some of them show an extended PSF and their FWHM$_{\rm WFPC2}$
values exceed our FWHM-threshold making them likely background galaxies.
The nature of the remaining very red point sources (indicated by squares
in Fig.~\ref{ps:sizes}), not rejected by these selection criteria but
only by the $V-K$ colour cut, is less clear. We discuss in turn some
possibilities of their nature:

{\it Foreground stars:} In the colour range $3.5\la V-K\la4.5$ mag and
the magnitude range $22.0\la V\la24.5$ mag we expect two stars in the
ISAAC field of view (6.25 arcmin$^2$), according to the model of
\cite{robin96}. The effective field of view is even smaller due to the
incomplete overlap between the ISAAC and WFPC2 field. That is,
significant stellar contamination at this red colours and faint
magnitudes can be considered as unlikely.

{\it Fuzzy red globular clusters:} In optical colours the extremely-red
objects are clustered around $V-I\approx1.4$ and 2.2 mag. Counterparts
of the redder objects could be the recently detected population of
extremely-red objects around NGC~1023, with typical colours
$V-I\approx2.2$, which appear to be more extended than the other
globular cluster candidates \citep{larsen00}. Using the resolution power
of HST, we measure the $\Delta(0.5-3)$ parameter \citep[which is simply
the magnitude difference between the flux within two different aperture
sizes; see e.g.][]{puzia99} to study relative sizes of our globular
cluster candidates. Figure \ref{ps:sizes} shows the relative size
distribution of globular cluster candidates and all other objects of our
samples. The wide-field chips of WFPC2 on-board HST resolve objects down
to 4.5 and 9.6 pc radius at the distance of NGC~3115 and NGC~4365,
respectively. Red objects, indicated by an open square (in
Fig.~\ref{ps:sizes}), were rejected by the $V-K$ colour selection only.

{\it Intermediate-age globular clusters:} We point out that the presence
of carbon-star dominated intermediate-age globular clusters, like they
are found in the LMC and SMC \citep{olszewski96}, is unlikely to explain
the extremely red colours. The strong IR flux of such AGB stars, yields
red $V-K$ colours but leaves the $V-I$ colour relatively unchanged. The
$V-I$ colours of the extremely-red objects are too red to be matched by
any simple stellar population at low redshift, regardless of its age and
AGB contribution.

{\it Background galaxies:} Unresolved background galaxies can, however,
also mimic a stellar PSF. In the case of our very red objects, we might
deal with the blue end of the known population of Extremely Red Objects
(EROs), identified as red-shifted and/or obscured galaxies at $z>0.85$
\citep[e.g.][]{liu00, daddi01}. The colour definition of EROs varies in
the literature and is typically around $R-K\ga4.5-6$ mag. The surface
density of EROs is a function of colour and limiting magnitude, and
increases from 0.07 arcmin$^{-2}$ for $R-K\geq6$ mag to 0.49
arcmin$^{-2}$ for $R-K\geq5$ mag for a sample with $K_{\rm lim}\leq19.0$
mag \citep{daddi01}. In our field of view, we expect 0.4 and 3.1 EROs,
respectively, down to $K=19$ mag. Our objects have similar colours as
the bluest EROs of which the distribution on the sky is inhomogeneous.
Hence, the number of very red objects in our field of view (two at
$K\sim19$ mag) is consistent with the surface density of EROs. The red
end of the NGC~4365 CMD shows also hints of an extremely-red population
of objects fainter than the photometric limit. All, except two sources
in this colour regime are rejected by the FWHM$_{\rm WFPC2}$ selection
and the completeness limit.

In conclusion, the nature of our extremely-red objects cannot be
definitely clarified from the photometric data only. Spectroscopy is
necessary to distinguish between cool M-type foreground stars, obscured
background galaxies, or (unobscured) extremely metal-rich young globular
clusters.

\subsection{Colour-Colour Diagrams}
\label{ln:colcol}
With the $VIK$ data in hand optical/near-IR colour-colour diagrams, such
as $V-I$ vs. $V-K$, can be constructed.

\begin{figure}
   \centering
   \includegraphics[width=8.5cm]{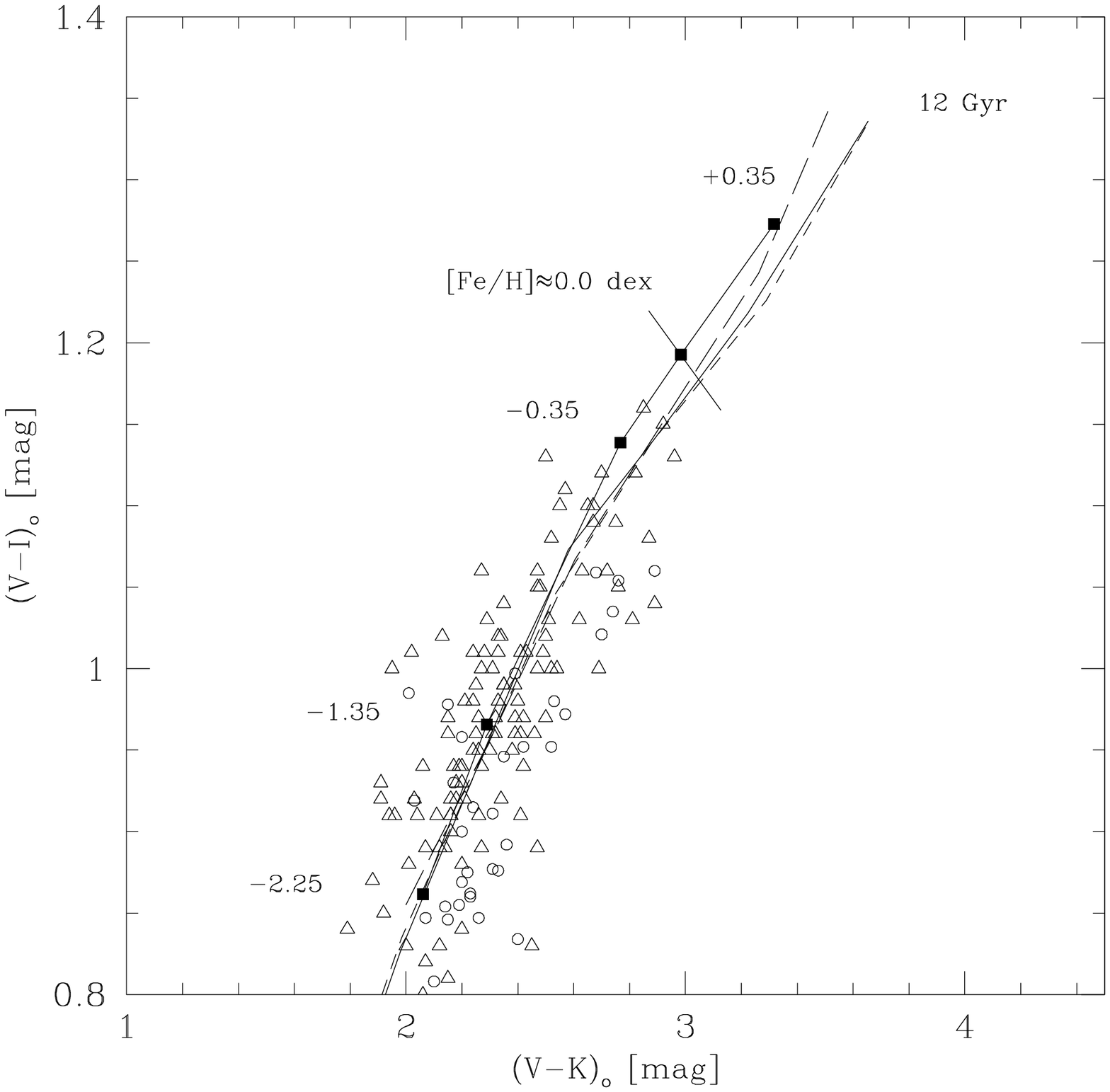}
      \caption{Variations of model predictions for a 12 Gyr old simple stellar
        population in the metallicity range [Fe/H]$=-2.25$ to $+0.35$
        dex. The lines indicate the predictions of \cite{maraston01}
        (solid), \cite{kurth99} (short-dashed), \cite{bc00} (dotted),
        and \cite{vazdekis99} (long-dashed). The metallicity on the
        iso-age line of \cite{maraston01} is labeled and indicated by
        filled squares. Open triangles are globular clusters of M31
        \citep{barmby00} while open circles mark the colours of Milky
        Way globulars \citep{aaronson02}. The $(V-I)$ data was taken
        from \cite{harris96}.}
         \label{ps:colcolmodall}
\end{figure}

In the following we compare globular cluster colours to different simple
stellar population (SSP) models and determine the mean age and
metallicity of the major sub-populations. We use the models of
\cite{vazdekis99}, \cite{kurth99}, \cite{bc00}, and \cite{maraston01}.
Absolute calibrations of SSP models are nowadays still uncertain.
However, the model-to-model differences of {\it relative} values are
small. This is particularly true for the age and metallicity ranges
covered by Milky Way and M31 globular clusters which serve as model
calibrators. Figure \ref{ps:colcolmodall} demonstrates model-to-model
variations for a stellar population of 12 Gyr with a metallicity range
$-1.7\la$[Fe/H]$\la+0.4$ dex. Also plotted are the globular clusters of
M31 and the Milky Way \citep{harris96,barmby00,aaronson02}. The plot
shows that at a fixed age, metallicity increases along the track towards
redder colours. Note, that there is very few data for M31 or Galactic
globular clusters in the solar-metallicity regime. Tracks of younger
ages fall generally below (i.e., have bluer $V-I$ colours than) older
tracks and the colour differences between tracks increases with younger
ages. For old ages the separation of tracks shrinks to values comparable
with photometric errors which diminishes the ability to distinguish
between ages.

\begin{figure*}
   \centering
   \includegraphics[width=10.5cm]{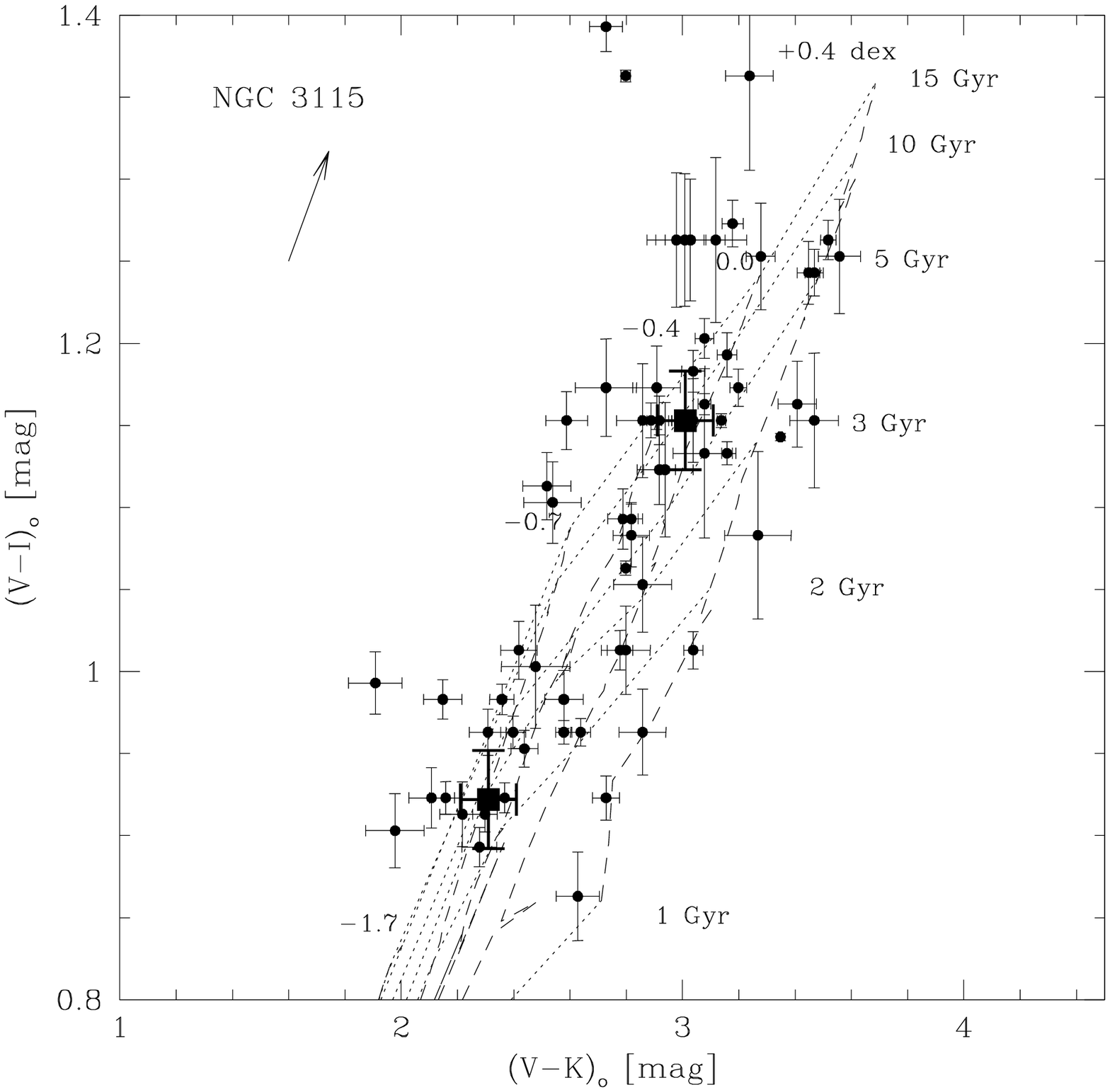}
   \includegraphics[width=10.5cm]{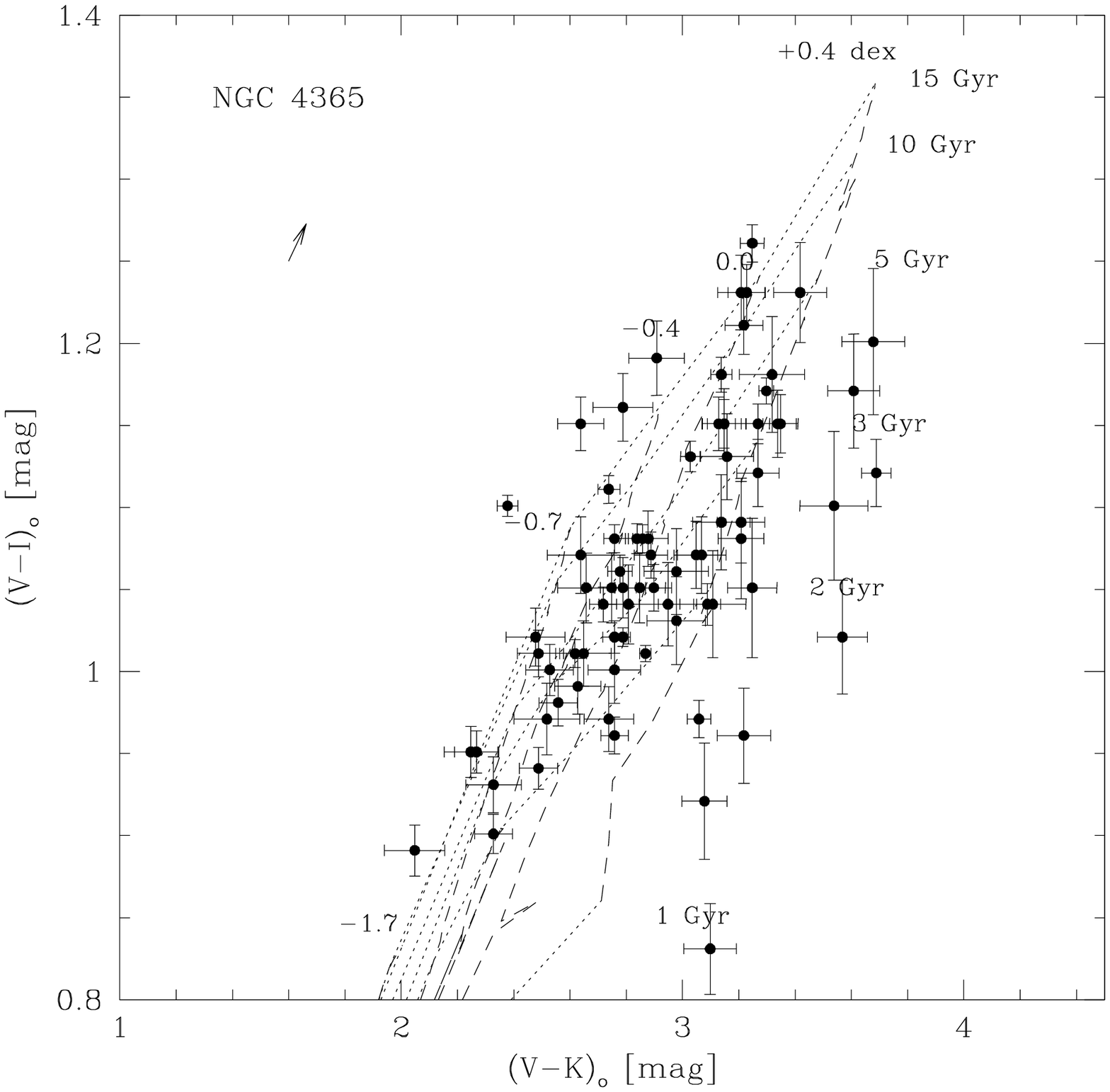}
      \caption{$V-I$ vs. $V-K$ colour-colour diagrams for NGC~3115 and
        NGC~4365. All data are reddening corrected according to
        Sect.\ref{ln:photcal} and \ref{ln:hstcal}. For comparison, we
        show six isochrones for simple stellar populations of 15, 10, 5,
        3, 2, and 1 Gyr (dotted lines from red to blue $V-I$ colours).
        The dashed lines represent iso-metallicity tracks changing from
        the lower left to the upper right from [Fe/H]$=-1.7$ to $+0.4$
        dex. The SSP model used here was adopted from \cite{bc00}. For
        NGC~3115, the large squares mark the mean colour of the modes as
        determined by KMM code (see Sect.~\ref{ln:kmm}). Arrows in the
        upper left corner of each panel indicate the applied extinction
        correction and the direction of reddening.}
         \label{ps:colcol}
\end{figure*}

\subsubsection{NGC~3115}

The colour-colour plot for NGC~3115 globular clusters in Figure
\ref{ps:colcol} show selected data with their photometric errors. Large
squares show the mean colours of the two sub-populations as determined
in the KMM analysis. Their error bars indicate the error of the mean as
defined by $\sigma_{\rm mean}=\sigma/\sqrt{N}$. As a guidance for the
eye, iso-age and iso-metallicity lines taken from the \cite{bc00} SSP
model are overplotted. Dotted lines show iso-age tracks for 1 to 15 Gyr
populations with metallicities $-1.7\leq$[Fe/H]$\leq +0.4$ dex.

Figure \ref{ps:colcol} reveals two main results. First, there is a broad
spread in metallicity with evidence for bimodality in the NGC~3115
cluster system. The globular cluster population can be split at around
$V-I\approx 1.05$ and $V-K\approx 2.6$ into two major sub-populations.
In general, the colour-colour diagram helps to pick out the
multi-modality which is not evident in $V-I$ or $V-K$ colour
distributions alone. The second result is that the two globular cluster
sub-populations have both old ages with a small or a negligible age
difference. Overall, the whole system seems to be dominated by old
globular clusters. The colour difference is, therefore, largely due to a
metallicity difference.

To quantify the age and metallicity difference between both globular
cluster sub-populations, we compare the difference in their $V-I$ and
$V-K$ colours to several SSP model predictions. We recall again that our
data sample is biased towards red globular clusters, i.e., we detect 1.5
times more red clusters in the NGC~3115 system than blue ones (see
Sect.~\ref{ln:completeness}). The blue peak of the $V-K$ colour
distribution might, therefore, vary as one goes deeper with the $K$ band
photometry, adding more blue globular clusters to the sample.

\begin{figure*}
   \centering
   \includegraphics[width=15cm]{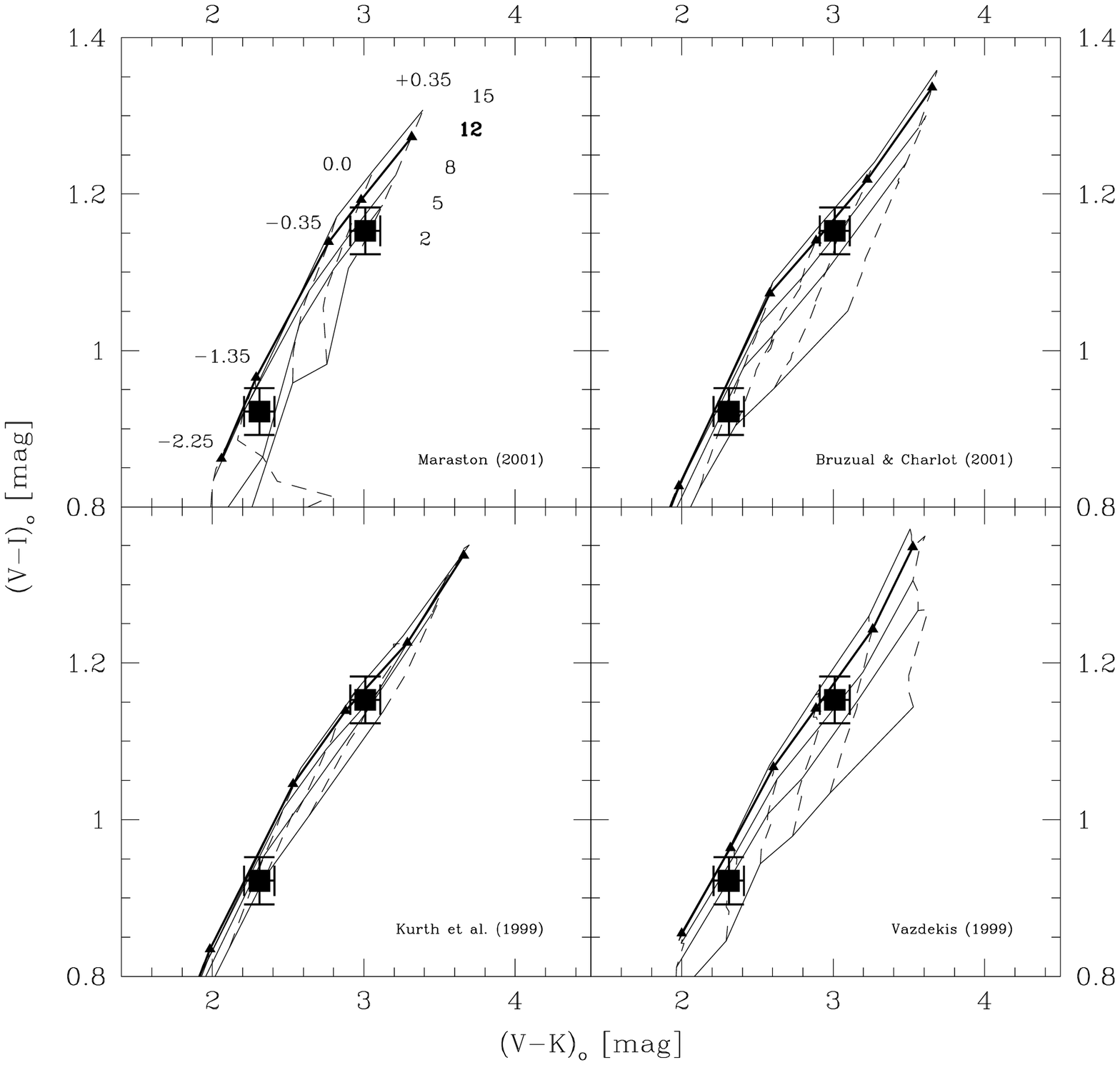}
      \caption{NGC~3115: $V-I$ vs. $V-K$ colour-colour diagrams for
        major globular cluster sub-populations. The four panels show the
        mean colour of the two major globular cluster sub-populations in
        NGC~3115 with four different SSP model grid.  In the upper left
        panel with the SSP model of \cite{maraston01} we labeled iso-age
        (solid lines) and iso-metallicity tracks (dashed lines) in the
        range $2-15$ Gyr and [Fe/H]=$-2.25$ to $+0.35$ dex. All other
        model grids have exactly the same iso-age lines. For better
        comparison we plotted the 12 Gyr iso-age line thicker. Triangles
        show the position of the labeled metallicity along the 12 Gyr
        age track.  The SSP models of \cite{bc00,kurth99,vazdekis99}
        have slightly different metallicity spacings. The are
        iso-metallicity tracks for [Fe/H]$=+0.4, 0.0, -0.4, -0.7, -1.7,$
        and $-2.3$ dex starting at red colours (upper right in each
        panel).}
         \label{ps:colcoldivmod}
\end{figure*}

Figure \ref{ps:colcoldivmod} shows the mean colours of the two
sub-populations overplotted on SSP model grids. From the model of
\cite{maraston01} we derive an age difference between the two
sub-populations of $\Delta t=(t_{\rm blue}-t_{\rm red})=2\pm4$ Gyr. The
other models yield age differences of $-5\pm6$ \citep{bc00}, $-4\pm5$
Gyr \citep{kurth99}, and $-4\pm5$ Gyr \citep{vazdekis99}. The mean age
difference is $\langle\Delta t\rangle=-2\pm3$ Gyr. The two globular
cluster sub-populations are coeval within 1$\sigma$.

With the assumption of coeval sub-populations we calculate the
metallicity difference which is the driving parameter for the $V-K$
colour offset. Applying the model of \cite{maraston01} yields
$\Delta$[Fe/H]$=|$[Fe/H]$_{\rm blue}-$[Fe/H]$_{\rm red}|=1.48\pm0.57$
dex. The models of \citeauthor{bc00}, \citeauthor{kurth99}, and
\citeauthor{vazdekis99} give $1.04\pm0.30$, $1.03\pm0.31$, and
$0.99\pm0.19$, respectively. The weighted mean metallicity difference is
$\Delta$[Fe/H]$=1.0\pm0.3$ dex. The metallicity difference derived with
the [Fe/H]--$(V-K)$ relation from Paper I (see Sect.~\ref{ln:coldistr})
is $1.28\pm0.32$ dex and is consistent within the errors with the model
predictions.

\subsubsection{NGC~4365}
\label{ln:colcoln4365}
The NGC~4365 colour-colour diagram is more complex than the one for the
NGC~3115 globular cluster system.

{\it Dust:} Can extinction by dust in NGC~4365 produce the colour
sequence we observe? The reddening vector is plotted in the upper left
corner of Figure \ref{ps:colcol}. If extinction by dust in NGC~4365
would play a role at all, one would expect this to have more impact
within the galaxy body than in the halo, which would cause the bulge
population to be systematically redder. Correcting for any additional
internal reddening would shift the clusters along the reddening vector
almost parallel to the iso-metallicity tracks towards younger ages (see
Fig.~\ref{ps:colcol}). However, there is no evidence for the existence
of significant amounts of dust in NGC 4365 \citep{jura87, goudfrooij94}.

{\it Intermediate-age clusters:} The most striking feature in Figure
\ref{ps:colcol} is a large population of globular clusters with
colours that are red in $V-K$ ($\sim3.0$) but intermediate in $V-I$
($\sim1.0-1.2$). According to SSP models this clusters fall in the
regime of $\sim Z_\odot-3Z_\odot$ metallicities and intermediate ages
of $\sim2-8$ Gyr. In general, the red globular clusters appear on
average younger than the blue sub-population.\footnote{We briefly
review results of previous GCLF studies which were based on optical
photometry only.  From recent HST photometry of 500 globular clusters
\cite{larsen01} find a difference in GCLF turn-over magnitudes $\Delta
m_{\rm TO}=0.8\pm0.4$ mag between the red and blue sub-population
which they cut at $V-I=1.05$ mag. With the red clusters being fainter,
their result implies that the red clusters are on average older and
more metal-rich than the blue sub-population, contrary to our
findings.  However, we note, that the $V-I$ colour distribution (see
Fig.~4 in \citealt{larsen01}) is very broad with no apparent
bimodality. Since \citeauthor{larsen01}'s $V-I$ separation splits
between blue and red clusters at roughly the colour of the
intermediate-age globular cluster population, their GCLF analysis
might be corrupted by the intermediate-age clusters that were not
accounted for.} With a few exceptions all metal-rich ([Fe/H]$\ga-0.4$
dex) clusters have $V-K$, $V-I$ colours indicative of intermediate to
young ages. We estimate this intermediate-age clusters to represent
$\sim40-80$\% of our observed globular cluster sample. This limits are
derived 1) by assuming all clusters red of the [Fe/H]$=-0.4$ dex
iso-metallicity line (see Fig.~\ref{ps:colcol}) to be intermediate-age
and 2) by taking into account the existing spread in colours of
globular clusters in Local Group galaxies and only counting objects
with colours redder than all Milky Way and M31 globular clusters and
with $V-K\ga3.0$ mag as intermediate-age clusters.

SSP models predict that the $V$ magnitude of a $\sim2-8$ Gyr globular
cluster is $\sim1.7$ to 0.6 mag brighter than that of a 15 Gyr old
cluster \citep[e.g.][]{bc00}. Thus we should be able to detect bright
globular clusters in the CMD. The NGC~4365 CMD (lower panel in
Fig.~\ref{ps:cmd}) shows indeed a population of relatively bright
globular clusters around $21\la V\la 22$ mag and $2.6\la V-K\la 3.4$
mag. Most of these objects, indeed, fall in the colour-colour diagram
(see Fig.~\ref{ps:colcol}) into the regime of metal-rich
intermediate-age populations with ages $\ga 8$ Gyr and metallicities
$Z\ga Z_\odot$. As a consistency check we, conversely, select clusters
in the colour-colour diagram (Fig.~\ref{ps:colcol}) with colours of
metal-rich intermediate-age populations (see colour definitions above)
and look for their $V$ magnitudes. They dominate the GCLF at brighter
magnitudes: At $V\la 22.0$ mag $\sim80$\% of all objects in the sample
are metal-rich and have intermediate ages. Unfortunately, the small size
and the colour bias of our sample does not allow to establish
statistically whether or not the GCLF of the metal-rich intermediate-age
population is brighter than that of the remaining globular clusters.

{\it Comparison of ages with known young clusters:} Clusters of pre-AGB
ages ($\la100$ Myr) would have not only different colours, but also
considerably brighter magnitudes than the old population and would thus
be easy to identify. A lower limit on the ages of the intermediate-age
population can be set by comparing their colours with the data for very
young ($\la1$ Gyr) globular clusters in NGC~7252 \citep{maraston01a}.
All NGC~7252 clusters show too blue $V-I$ colours compared with the bulk
of intermediate-age clusters in NGC~4365. This implies that the latter
are at least 1 Gyr old.

A recent study of globular clusters in NGC~1316 revealed an
intermediate-age ($3\pm0.5$ Gyr), solar-metallicity globular cluster
sub-population \citep{goudfrooij01a, goudfrooij01b}. The brightest and
not-reddened intermediate-age cluster candidates have mean colours
$V-I\approx1.05$ and $V-K\approx2.87$. This colours, compared with the
mean locus of globular clusters in NGC~4365 (cf. Fig.~\ref{ps:colcol}),
indicate that some of the intermediate-age globular clusters in NGC~4365
have counterparts in NGC~1316 which are $\sim3$ Gyr old.

{\it Spatial distribution:} Interestingly enough, the young globular
clusters with colours $V-K\ga 2.6$ seem to form a 'bulge' or 'disk'
population, as the distribution of those clusters is clearly elongated
along the major axis of the galaxy body (cf.\ Fig.~\ref{ps:fov}). The
major axis of the disk aligns with the major axis of the galaxy body. In
contrast, blue globular clusters form a halo population around NGC~4365.
Unfortunately, our field of view is concentrated to the south-east only
and does not allow to check for consistency on the opposite side
(north-west) of the major axis.

\subsection{Aging the NGC~4365 Globular Cluster System}
\label{ln:aging}

\begin{figure*}
   \centering
   \includegraphics[width=16cm]{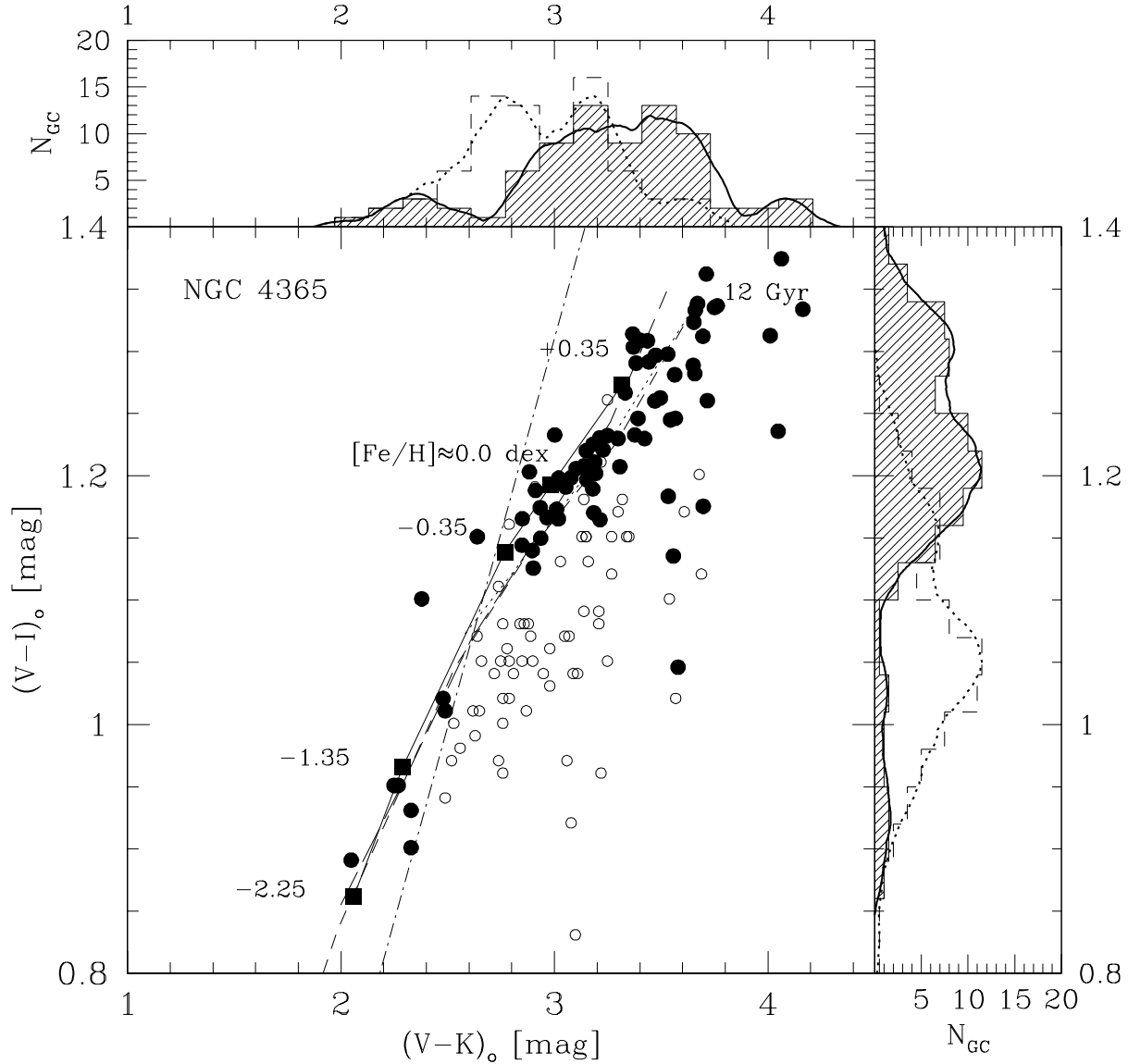}
      \caption{NGC~4365: Colour-colour diagram for the evolved globular
        cluster system. Solid circles show the evolved data while open
        circles indicate the colours of the observed data set.  Since an
        average evolution calculated from four SSP models was adopted we
        plot 12 Gyr iso-age tracks for each model with metallicities
        $-2.25<$[Fe/H]$<0.35$ dex. The metallicity sequence is indicated
        by solid squares. We used models from \cite{maraston01} (solid),
        \cite{kurth99} (short-dashed), \cite{bc00} (dotted), and
        \cite{vazdekis99} (long-dashed). All clusters right of the
        dot-dashed line, which is the limit of the applied linear
        interpolation, were aged.}
         \label{ps:n4365aged}
\end{figure*}

In the following we let the entire globular cluster system in NGC~4365
age for 10 Gyr and study their evolution in the colour-colour diagram.
We assume a passive colour evolution due to age only. Each single
globular cluster is aged individually for 10 Gyr.

Following \cite{buzzoni95} a detailed inspection of each model shows
that the parameter range $5\la t\la15$ Gyr and $-0.7\la$[Fe/H]$\la0.4$
dex (except for \citeauthor{maraston01} models, for which we use a
smaller metallicity range $-0.35\la$[Fe/H]$\la0.35$ dex) allows a simple
parameterization of colours by age and metallicity. The parameterization
can be written as a first approximation:
\begin{eqnarray}
\label{eqn:param}
\Delta Colour_i = \alpha_i \Delta log ~t + \beta_i \Delta {\rm [Fe/H]}
\end{eqnarray}
where the slopes $\alpha_i, ~\beta_i$ are determined from the SSP models
and are given in Table \ref{tab:slopes}.
\begin{table*}[!ht]
\centering
\caption[width=\textwidth]{Slopes $\alpha_i, ~\beta_i$ as determined
  from the SSP models. The mean $\alpha_{V-K}$ was calculated without
  the \citeauthor{vazdekis99} value.}
\label{tab:slopes}
\begin{tabular}{l l c c r}
\hline
Model & $i$ & $\alpha_i=\partial /\partial(log~t)$&$\beta_i=\partial
/\partial$[Fe/H] & parameter space \\
\noalign{\smallskip}
\hline
\noalign{\smallskip}
\cite{bc00}       
   & $V-I$ & $0.239\pm0.029$ & $0.236\pm0.006$ & $5<t<15$ Gyr \\
   & $V-K$ & $0.459\pm0.150$ & $0.964\pm0.015$ & $-0.7<$[Fe/H]$<+0.4$ \\
\noalign{\smallskip}
\cite{kurth99}
   & $V-I$ & $0.218\pm0.043$ & $0.259\pm0.011$ & $5<t<15$ Gyr \\
   & $V-K$ & $0.529\pm0.092$ & $0.981\pm0.026$ & $-0.7<$[Fe/H]$<+0.4$ \\
\noalign{\smallskip}
\cite{maraston01}
   & $V-I$ & $0.260\pm0.020$ & $0.207\pm0.011$ & $5<t<15$ Gyr \\
   & $V-K$ & $0.554\pm0.017$ & $0.825\pm0.020$ & $-0.35<$[Fe/H]$<+0.35$ \\
\noalign{\smallskip}
\cite{vazdekis99}
   & $V-I$ & $0.201\pm0.052$ & $0.265\pm0.006$ & $5<t<15$ Gyr \\
   & $V-K$ & $0.065\pm0.133$ & $0.862\pm0.030$ & $-0.7<$[Fe/H]$<+0.4$\\
\noalign{\smallskip}
\hline
\noalign{\smallskip}
$\langle$mean$\rangle$
   & $V-I$ & $0.230\pm0.026$ & $0.242\pm0.026$ & $5<t<15$ Gyr \\
   & $V-K$ & $0.514\pm0.040$ & $0.908\pm0.076$ & $-0.35<$[Fe/H]$<+0.35$ \\
\noalign{\smallskip}
\hline
\end{tabular}
\end{table*}
We use linear interpolation in each model grid and average the
obtained ages and metallicities from all four models. For clusters
with redder colours than the model parameter space (see Table
\ref{tab:slopes}) we use an extrapolation to high-metallicities, being
aware of the insecure prediction power of SSP models in this range.
Clusters which have bluer colours than the model parameter space are
dropped from the evolution due to apparent non-linearities in the
grids beyond [Fe/H]$\la -0.7$ dex and an expected small colour
evolution. Since we assume passive evolution one can write
\begin{eqnarray}
\label{eqn:evol}
\Delta(V-I) =& 0.230\log(1+\frac{10 {\rm Gyr}}{t})\\
\Delta(V-K) =& 0.514\log(1+\frac{10 {\rm Gyr}}{t})
\end{eqnarray}
where $t$ is the current age of each single globular cluster
determined from the colour-colour diagram and $\Delta(V-I)$ and
$\Delta(V-K)$ are the colour changes over a 10 Gyr lifetime. The
mathematical form of these equations implies a rapid evolution in
colour for intermediate-age globular cluster. As the clusters get
older, the colour evolution slows down, i.e. old clusters are hardly
affected in colour by the aging procedure.

We emphasize again that both upper equations are only valid for stellar
populations older than $\sim5$ Gyr (see Table~\ref{tab:slopes}) and
that, especially for younger ages, the relations represent insecure
extrapolations. Figure \ref{ps:n4365aged} shows the evolution of the
colour-colour diagram and the according colour distributions.

Only clusters which fall right to the dot-dashed line are considered in
the aging process. This boundary is defined by the limitation of our
parameter space (see Table~\ref{tab:slopes}). Moreover, there are very
few objects left of the line for which the models predict almost no
colour evolution. The aging shows that the 10 Gyr older globular cluster
system (solid dots) contains a significant fraction of clusters with
$Z>Z_\odot$ metallicities. Clusters with extremely red colours $V-K>4$,
which correspond to metallicities $Z\ga 10Z_\odot$, are present in the
new colour-colour distribution. Note, however that these extreme colours
are the predictions of the extrapolated parameter range and might change
when detailed models of extremely metal-rich stellar populations are
available.

The aged $V-I$ and $V-K$ colour distributions (thick lines
Fig.~\ref{ps:n4365aged}) show a very broad flat-top red peak. This broad
peak suggests a wider metallicity dispersion than seen in the metal-rich
sub-system of Galactic globular clusters, which has a width of about 0.3
dex. This metallicity width translates into a scatter of $\sim0.25$ mag
in $V-K$ and $\sim0.15$ mag in $V-I$. The red peak of the evolved
intermediate-age clusters in NGC~4365 is at least twice as broad.
However, both colour distributions, although statistically
insignificant, show hints for a substructure of the red peak which might
consist of two or more red sub-populations (see the $V-K$ colour
histogram in Fig.~\ref{ps:n4365aged}).

The intermediate-age sub-population of today's NGC~4365 globular
clusters will mature to colours which are comparable to colours of the
red (metal-rich) sub-population in M87 (Paper I). These broad red
sub-populations are also found in other giant elliptical galaxies, such
as NGC~4472 \citep[e.g.][]{geisler96,puzia99}, NGC~3585, NGC~4526,
NGC~4649, NGC~5846 \citep[e.g.][]{gebhardt99, larsen01, kundu01a,
  kundu01b}.

\subsection{Comparison with the Integrated Light of the Host}
\label{ln:colcolhostcomp}
Figure \ref{ps:mwm31colcol} shows the $V-I$ and $V-K$ colours of the
integrated light of NGC~3115 and NGC~4365 along with the colours of
their globular cluster systems. The effective $V-I$ colour was taken
from \cite{buta95}. The $V-K$ colour was measured by \cite{frogel78} in
different apertures which are not the effective apertures. We use the
colour given by the aperture which is closest to the effective radius
and calculate the expected change to an aperture of effective radius.
All values are documented in Table \ref{tab:galdat}.

From the comparison with SSP models both galaxies show diffuse stellar
light colours which are consistent with a $\sim12$ Gyr old $Z>Z_\odot$
metallicity stellar population. However, the observed colours are
luminosity-weighted averages and can be reproduced by many different
mixes of stellar populations with a wide range of ages and
metallicities. The major difference between the two galaxies appears
when the stellar light is compared with their globular cluster systems
(see Fig.~\ref{ps:mwm31colcol}). The globular clusters in NGC~3115
seem to have the same age as the galaxy body. The colours of the
integrated light fall onto the metallicity sequence which is set up by
the globular clusters. A few very red globular clusters share the
colours of the stellar light. This clusters appear to be associated
with the stellar disk of NGC~3115.

In NGC~4365, the globular cluster system shows features that are not
readily apparent in the diffuse stellar light. The galaxy's average
stellar population is metal-rich and relatively old and the clusters
span a wide range of metallicities and ages.

The triaxial galaxy NGC~4365 hosts a decoupled core which rotates around
the minor-axis \citep{wagner88,bender92,surma95} while the main body
rolls around its major axis. The first detailed abundance study found
the innermost core stellar population being metal-rich
($\ga2.3Z_\odot$), relatively young ($\la7\pm1.5$ Gyr) and excessively
enriched in $\alpha$-elements \citep{surma95}. \cite{davies01} performed
a subsequent integral-field spectroscopic study covering the galaxy out
to $\sim3$ kpc and found the same age ($\sim14$ Gyr) for the two
distinct components, solar metallicity and a rather mild
$\alpha$-element enhancement of the core. However, their Figure 2 shows
that there is an age gradient in the very center ($r\leq1.6$\arcsec )
where the innermost population is $\sim12$ Gyr old. The quite
significant difference in age between the former two studies can be
attributed to the use of two different SSP models.

The index measurements of both studies for H$\beta$ and MgI are
consistent with each other. In particular, while \citeauthor{surma95}
use \cite{worthey92} models, \cite{davies01} use the models of
\citep{vazdekis99} which, in contrast to Worthey's models, include late
evolutionary phases, such as the HB (horizontal branch), AGB, EAGB, and
PAGB. Since the HB can significantly contribute to the H$\beta$ flux
\citep{maraston00}, which is commonly used as an age-sensitive index,
the predictions of a model which includes the HB will yield higher ages
for same H$\beta$ measurements as models without HB modeling. The
difference in age between the previous two studies can therefore be
fully understood from the choice of models.

The latest Vazdekis99 models are also used in this study. Therefore, the
spectroscopic and photometric predictions for age and metallicity can
directly be compared with each other. For NGC~4365, with the photometric
$VIK$ values of Table \ref{tab:galdat} we obtain an age of $12\pm4$ Gyr
and roughly solar metallicity. Considering that the photometric values
are an average over a large radial range, the predictions of photometric
and spectroscopic parameters from the Vazdekis99 models are consistent
with each other.

\subsection{Comparison with the Milky Way and M31 Globular Clusters}
\label{ln:mwm31comp}
\begin{figure*}
   \centering
   \includegraphics[width=8.5cm]{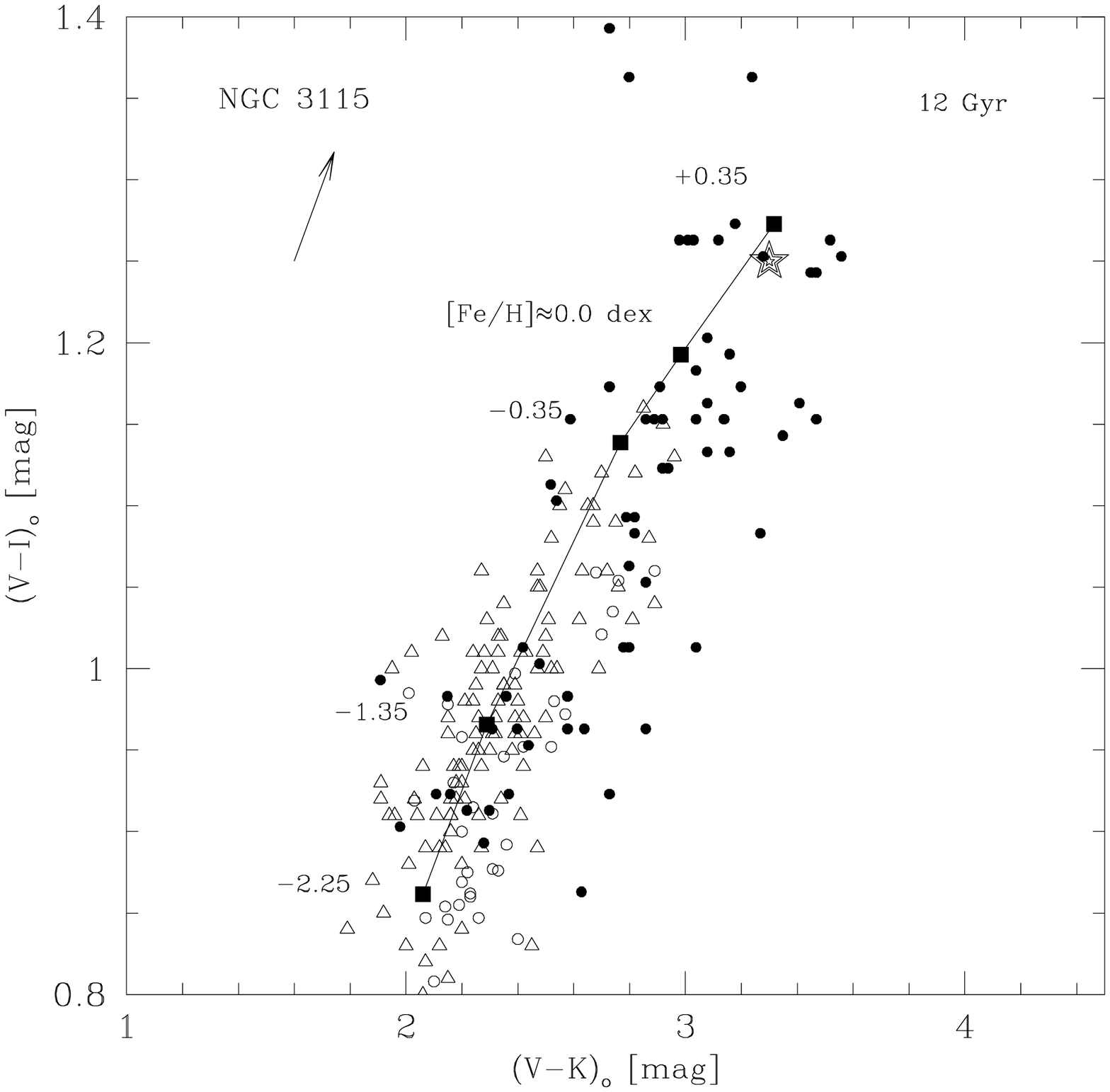}
   \includegraphics[width=8.5cm]{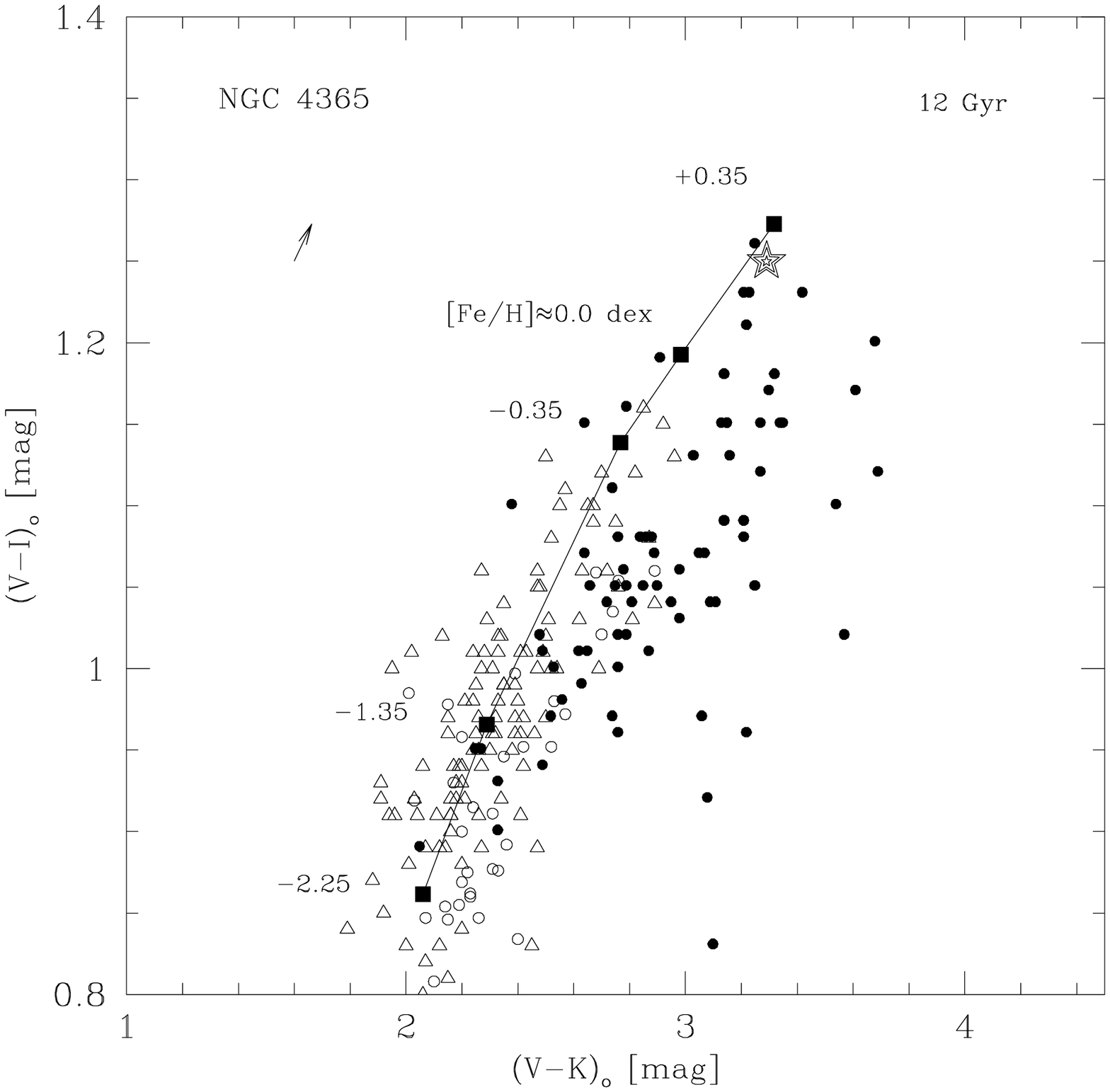}
      \caption{Comparison of globular clusters in NGC~3115 (left panel)
        and NGC~4365 (right panel) with globular clusters in the Milky
        Way and M31. Milky Way and M31 clusters are shown by open
        circles and open triangles. The stars show the colour of the
        diffuse stellar light (see Table~\ref{tab:galdat}). As a guide
        to the eye, we plot a 12 Gyr iso-age line for different
        metallicities in the range $-2.25\leq$[Fe/H]$\leq +0.35$ dex
        taken from \cite{maraston01}. The metallicity is increasing to
        redder colours and is indicated by solid squares. Note, that in
        the two Local Group spirals the high-metallicity range contains
        only few globular clusters which are, moreover, prone to have
        inaccurate reddening corrections.}
\label{ps:mwm31colcol}
\end{figure*}

We compare the colour-colour diagrams of NGC~3115 and NGC~4365 clusters
with the ones of Milky Way and M31 globular clusters. Figure
\ref{ps:mwm31colcol} shows the comparison of the colour-colour diagrams.
The globular cluster system of NGC~3115 compares well with the globular
cluster system of either Local Group spiral. The globular cluster
systems of NGC~3115, Milky Way, and M31 appear to be on average old and
roughly coeval.

The few detected blue clusters in NGC~4365 show good agreement with the
mean locus of metal-poor clusters in M31 and the Milky Way. The NGC~4365
globular cluster system extends also to much higher $V-K$ at similar
$V-I$ colours than the Local Group systems, implying high metallicities
and intermediate to young ages (see also Fig.~\ref{ps:colcol}).

\section{Discussion}
\label{ln:discussion}
\subsection{NGC~4365, a Cooling-Down Galaxy}
\cite{davies01} find a luminosity-weighted age of $\sim12$ Gyr for the
solar-metallicity kinematically-decoupled core (KDC) in NGC~4365 using
the diagnostic plot (H$\beta$ vs.~[MgFe]) and SSP models of
\cite{vazdekis99}. From the comparison of our near-IR/optical colours
with the same models, we find that $\sim40-80$\% of our sample are
metal-rich ($Z\ga0.5\, Z_\odot$) globular clusters with ages between
$\sim2-8$ Gyr. However, within our observational measurement errors
the age distribution, derived from the linear interpolation of SSP
models, is consistent with a single-burst of globular-cluster
formation which took place $\sim3-4$ Gyr ago. Together with the
metal-rich globular clusters a part of the stellar population in the
KDC must have been formed from the same pre-enriched gas in a
significant star-formation event. In fact, \cite{davies01} find that
the $\sim2$ Gyr younger luminosity-weighted age of the KDC (compared
to the outer parts of the galaxy) can be accounted for by a
contaminating, 5 Gyr old stellar population of solar metallicity. This
young population would amount 6\% of the mass inside $r\la1.6$\arcsec\
or $\la160$ pc.

It is tempting to associate the intermediate-age globular cluster
population with the kinematically decoupled core. Within the
prediction accuracy of SSP models this young KDC population is
consistent with being the diffuse counterpart of the stellar
population in metal-rich intermediate-age globular clusters. Numerical
simulations of gaseous mergers \cite[see e.g.][for a review]{barnes92}
predict the settling of the progenitors' gas into the inner parts of
the merger remnant within $\sim10^8$ years. If globular clusters and
the central stellar population were in fact formed from the same gas,
both stellar populations have to be almost coeval.

We can compare the stellar mass contained in an intermediate-age
component in NGC~4365 with the mass of the intermediate-age globular
cluster population to test how consistent these are with forming in
the same event. To calculate the mass of the diffuse stellar
component, we begin with the decoupled core. \citeauthor{davies01}
found that the intermediate-age component makes up 6\% of the stellar
mass of the decoupled core. Given the total stellar mass estimate for
the core of $M_{\rm core}\la (8.1\pm2.5)\cdot10^9 M_\odot$ from
\cite{surma95}, the resulting mass in the core of the intermediate-age
diffuse stellar component is $(4.9\pm2.5)\cdot10^8 M_\odot$.

This mass is a lower limit to the total mass of the diffuse
intermediate-age stellar population because it is unlikely that this
component is found only in the very central region. In particular, the
intermediate-age globular clusters are observed over a fairly extended
region, and it would be difficult to arrange for the corresponding
diffuse component not to have a similar extended spatial distribution.
Therefore, in order to estimate the total stellar mass of the galaxy
at intermediate ages, we need to estimate the fraction of the total
galaxy mass coming from the intermediate age population. This fraction
can not be greater than a few percent, or the extended
intermediate-age population would have been detected by
\citeauthor{davies01} as they detected such a component in the central
region. Conversely, the mass fraction of the intermediate-age
component must be at least a few-tenths of a percent based on
comparing the intermediate-age stellar mass of $(4.9\pm2.5)\cdot10^8
M_\odot$ observed in the core to the total galaxy stellar mass of
$\sim4.8\cdot10^{11}M_\odot$ \citep{bender89}.

To bracket these possibilities, we carry out a representative
calculation with a 1\% fraction of the overall stellar mass of the
galaxy in an intermediate-age component, and show how the results are
affected by varying this in either direction. This fraction gives a
stellar mass over the whole galaxy in the intermediate-age component
of $\sim5.2\cdot10^9M_\odot$.

What is the corresponding mass currently in the intermediate-age GC
system? After correcting our sample for incompleteness in luminosity
and spatial coverage and with the assumption that $\sim40-80$\% of our
sample globular clusters have intermediate-ages, we find that the
total number of intermediate-age globular clusters in NGC~4365 is
$\sim 150$ (within a factor of 2). To determine the mass in this
system, we adopt the two-component mass function of
\cite{mclaughlin96} over the mass range of $10^4-10^8 M_\odot$,
appropriate for the GCs of giant elliptical galaxies. This gives a
total mass of $\sim1.6\cdot10^8M_\odot$ in the globular cluster
system. A similar result is found by simply counting up the estimated
mass of the observed clusters and accounting for the spatial areas not
covered by our observations.

The efficiency of formation and survival to several Gyr of the
intermediate-age globular cluster system is 3\%. This follows directly
from the derived mass of $\sim1.6\cdot10^8M_\odot$ for the
intermediate-age globular cluster system and the corresponding
$\sim5.2\cdot10^9M_\odot$ for the diffuse stellar component of the
intermediate-age system. If the adopted mass fraction of the galaxy is
too high, the formation and survival efficiency increases to rather
high values of 10\% or more. If the adopted mass fraction of the
galaxy in intermediate-age component is a factor of a few too low,
this number drops to about 1\%. It is difficult to make the data
consistent with a formation efficiency less than 1\%.

How do these formation and survival efficiencies compare to other
observations? Observations of strongly starbursting systems show that
$\sim$ 20\% of the stars form in compact star clusters
\citep[e.g.][]{meurer95, whitmore99, larsen99, zepf99}. This
represents the peak formation efficiency of globular clusters. Studies
of older mergers find that a few percent of the stars are in globular
clusters \citep[e.g.][]{schweizer96, miller97, zepf99,
goudfrooij01b}. The best determined case for older globular cluster
systems is probably the Galactic halo, in which about 1\% of the stars
are in globular clusters. A global average of all spheroidal galaxy
systems is around 0.3\% \citep[e.g.][]{mclaughlin99}. However, there
are large variations between galaxies, and within sub-populations in
individual galaxies. Dynamical evolution may significantly reduce the
number of globular clusters observed at old ages compared to initial
values at young ages \citep[e.g.][]{gnedin97, vesperini00}.

The comparison of the formation and survival efficiency of the
NGC~4365 intermediate-age globular cluster system to other systems
suggests that NGC~4365 and older mergers have about 2\% of their stars
in globular clusters, with an uncertainty of a factor of a few. It is
possible that the stellar mass attributed to the younger component
should be restricted to that in the core alone, in which case the
formation and survival efficiency of the NGC~4365 globular cluster
system would be more like the peak efficiency seen in recent mergers
with little subsequent dynamical evolution. Underestimating the mass
in the diffuse intermediate-age population would reduce the formation
and survival efficiency by a factor of a few, but it is difficult to
make the data consistent with globular cluster formation efficiencies
less than a percent or so.

We can also compare the metallicity of the intermediate-age globular
cluster and diffuse stellar components. NGC~4365 shows
super-metal-rich globular clusters ($Z>2\,Z_\odot$) while the
luminosity-weighted stellar light of the decoupled core reveals a
roughly solar metallicity \citep{surma95, davies01}. This is likely
due to a mixing effect of the integrated light of stellar populations
in the core. If stars and globular clusters were formed from the same
pre-enriched material, super-solar metallicities are expected in the
core, as well. The small fraction of light from super-metal-rich stars
is likely to be washed out by insufficient resolution of the inner
structures. With a spatial resolution of $\sim1.4$\arcsec\ (radial
size of the decoupled core $\sim8$\arcsec ) \cite{surma95} report that
the metallicity is at least $2.3Z_\odot$ in the center of NGC~4365. At
face value, the core metallicity and the metallicity of the most
metal-rich globular clusters are consistent.

\subsection{NGC~3115, a Quiescent Galaxy}
It seems that the colour difference between the two major
sub-populations of globular clusters in NGC~3115 is mainly due to a
difference in metallicity, rather than in age. Within the errors this
result is still consistent with the findings of \cite{kundu98} who
report, using optical photometry only, that the metal-rich clusters are
about $\sim3$ Gyr younger than the metal-poor ones.

In contrast to NGC~4365, NGC~3115 seems to have experienced no major
star-formation event in the last $\sim$10 Gyr. Both its globular
cluster sub-populations are roughly that old or older. This resembles
the situation in the globular cluster system of the Milky Way, where
the two sub-systems, corresponding to thick-disk/bulge and halo
populations, differ mainly in metallicity \citep[][etc.]{minniti95,
cote00} with a relatively small age difference where the metal-rich
clusters may be slightly younger than their metal-poor counterparts
\citep{rosenberg99}.

Quantitatively, in NGC~3115 it remains to be seen whether or not the
specific frequency $S_{\rm N}$ (number of globular clusters per unit
luminosity, see \citealt{harris81}) for the metal-poor stellar
populations (i.e., the ratio of metal-poor globular clusters to
metal-poor halo stars) is also roughly three times higher than for the
metal-rich stellar populations, as it is observed in NGC~5128 and M31.

Given the limited information about the halo stars, we conclude that
the globular cluster system and the majority of halo stars in NGC~3115
most likely formed in two epochs and/or by two different mechanisms
more than $\sim$10 Gyr ago. Ever since then, the stellar populations
in globular clusters and in the galaxy halo seem to have undergone a
purely passive evolution.

\section{Summary}

Using accurate optical and near infrared photometry of the globular
cluster systems in NGC~3115 and NGC~4365 we found the following:

\begin{itemize}
  
\item The large color baseline introduced by the new $K$-band data
  allowed us to find, for the first time, a very metal-rich
  intermediate-age population of globular clusters in NGC~4365. The
  clusters have ages between $\sim2-8$ Gyr and metallicities $\sim
  0.5Z_\odot-3Z_\odot$. They form a flattened population, which is aligned
  with the galaxy's major-axis, while the metal-poor clusters populate
  the halo. We emphasize that such intermediate-age populations can only
  be found from photometric data by obtaining deep $K$-band data in
  addition to high-quality optical imaging from HST/WFPC2.
  
\item After $\sim$10 Gyr of passive evolution these globular clusters
  will show colours similar to super metal-rich globular clusters in
  giant elliptical cluster galaxies, such as M87.
  
\item NGC~3115 hosts an old and bimodal globular cluster population.
  From comparison to four different SSP models we found the blue and the
  red globular cluster sub-populations being coeval to within the errors
  ($\pm3$ Gyr).
  
\item We determine a metallicity difference between the red (metal-rich)
  and blue (metal-poor) globular cluster sub-population in NGC~3115 as
  $\Delta$[Fe/H]$=1.0\pm0.3$ dex. Thus, for this particular globular
  cluster system, metallicity is primarily responsible for the
  bimodality.
  
\end{itemize}
 
\begin{acknowledgements}
  We would like to thank the ESO user support group and the ESO
  science operations group for having carried out very successfully
  our programme in service mode. We thank Judy Cohen for providing
  near-IR data for Milky Way globular clusters. We are also grateful
  to Stephane Charlot and Claudia Maraston for providing their stellar
  population models prior to publication. THP gratefully acknowledges
  the financial support during his visits at Yale University where
  parts of this work were performed. THP also acknowledges the support
  by the German \emph{Deut\-sche For\-schungs\-ge\-mein\-schaft,
    DFG\/} project number Be~1091/10--1. SEZ acknowledges support for
  this work from NASA LTSA grant NAG5-11319. MH acknowledges support
  through Proyecto Fondecyt 3980032. DM is supported by FONDAP Center
  for Astrophysics and FONDECYT. MH, DM, and PG thank the European
  Southern Observatory for supporting visits during which part of this
  work was carried out. PG was affiliated with the Astrophysics
  Division of the Space Science Department of the European Space
  Agency during part of this project.
\end{acknowledgements}

\bibliographystyle{apj}

\end{document}